Linking theory and empirics: a general framework to model opinion formation processes

Ivan V. Kozitsin[1, 2]

[1]*Laboratory of Active Systems, V. A. Trapeznikov Institute of Control Sciences of Russian Academy of Sciences, 65 Profsoyuznaya street, Moscow, 117997, Russian Federation*

[2]*Department of Higher Mathematics, Moscow Institute of Physics and Technology, 9 Institutskiy per., Dolgoprudny, Moscow Region, 141701, Russian Federation*

**Email**: kozitsin.ivan@mail.ru

We introduce a minimal opinion formation model, which is quite flexible and can reproduce a broad variety of the existing micro-influence assumptions and models. At the same time, the model can be easily calibrated on real data, upon which it imposes only a few requirements. From this perspective, our model can be considered as a bridge, connecting theoretical studies on opinion formation models and empirical research on social dynamics. We investigate the model analytically by using mean-field approximation and numerically. Our analysis is exemplified by recently reported empirical data drawn from an online social network. Employing these data for the model's parameter calibration, we demonstrate that the model may reproduce fragmented and polarizing social systems. Furthermore, we manage to generate an artificial society that features properties quantitatively and qualitatively similar to those observed empirically at the macro scale. This ability became possible after we had advanced the model with two important communication features: selectivity and personalization algorithms.

## 1. Introduction

Models of opinion dynamics (aka social-influence models) concern how individuals change their opinions as a response to information their receive from social environments. Understanding the processes of opinion dynamics is important due to its applications in many fields, including policy-making, business, and marketing. This branch of modeling is naturally interdisciplinary, attracting scholars from different fields, including social psychology, control theory, and physics. Despite the theoretical side of these models having been seriously advanced, the problem of its applications to describing real social processes remains an important question (Castellano et al., 2009; Flache et al., 2017; Mäs, 2019; Proskurnikov & Tempo, 2017, 2018). This issue is rooted in the complex nature of the social systems. To be more precise, it is extremely difficult to calibrate parameters of the underlying social-influence models, which operate with hardly formalizable entities. One prominent example is that of opinions themselves, which are intrinsic components of such models (Mäs, 2019).

The proliferation of online social networks (OSNs) has made it possible to identify the dynamics of users' opinions on a large scale by applying machine learning techniques (Barberá, 2014; Chang et al., 2017). Combining these methods with tools elaborated in the field of social network analysis, one can obtain both dynamics of opinions and information on social connections between individuals (M. E. J. Newman, 2018b, 2018a). Further, recent research has proposed a methodology to identify not only the structure of ties but also their weights, which describe how well these ties conduct social influence (aka influence networks) (Ravazzi et al., 2021). This information may be effectively integrated into the existing social-influence models, allowing to calibrate them, validate, and, further, make necessary predictions.

However, the step involving the integration of information gathered from OSN into the models is hampered because there is a substantial scope of the opinion dynamics models, and each model may require a specific data format. Hence, empirical data gathered through an experiment (most likely an expensive and time-consuming one) may be useful in one case but, unfortunately, unacceptable or requiring a lot of extra work with data in other situations. Thus, each new dataset on opinion dynamics will potentially have a limited area of application in the sense that it could be investigated only by a restricted number of opinion formation models.

Therefore, we propose a quite general and, nonetheless, minimal model of opinion formation. On the one hand, the model is extremely flexible and can approximate a broad variety of the existing micro-influence assumptions and models. On the other hand, our model can be easily calibrated on empirical data, upon which it imposes a relatively small number of requirements. From this perspective, this model can serve as a bridge, connecting theoretical studies on opinion formation models on the one hand and empirical research on the other. We investigate the model analytically and numerically and exemplify our analysis with recently reported empirical data drawn from an online social network.

The remainder of this paper is organized as follows. Section 2 reviews the related literature. In Section 3, we elaborate upon the model and discuss it. Section 4 presents the analytical results obtained from use of the model. In Section 5, we describe the design of numerical experiments that we use to investigate the model's behavior. Section 6 presents the results of the numerical experiments, and in Section 7, we discuss them. Concluding remarks are provided in Section 8, and, finally, the Appendix includes supplemental information.

## 2. Literature

Social-influence models are numerous, and it is extremely difficult to classify all of them correctly. However, most of them can be grouped into main classes to some extent. Within this paper, we concentrate on the micro-level models, whereby the opinions of every individual are initialized and can both change and cause changes. In such models, the modeler analyzes how influence processes at the micro scale affect the resulting state of the social system at the macro scale. In contrast, so-called macro-level models describe the behavior of macroscopic variables, such as populations of individuals espousing particular opinions (Rashevsky, 1939).

The literature emphasizes three main classification criteria. (1) Time in the model: continuous (Abelson, 1964) or discrete (DeGroot, 1974); (2) opinions: continuous (Friedkin & Johnsen, 1990) or discrete (Moretti et al., 2013; Sznajd-Weron & Sznajd, 2000); and (3) micro-assumption of social influence (see below). Because the empirical data are typically gathered in discrete time, we will focus hereafter on discrete-time models. Besides, we will assume that opinions are represented by scalar quantities describing individuals' positions on a single issue. More complex situations could arise when several topics (logically connected or independent) are analyzed at once (Friedkin et al., 2016). However, gathering data on individuals' opinions on two or more topics simultaneously is a challenging and costly task. On this basis, we will focus on scalar opinions.

There are three main micro-assumptions regarding social influence, built upon a continuous opinion space (Flache et al., 2017). In that space, in the case of one-dimensional opinions, one can say that an individual's opinion is affected by positive (aka assimilative) influence from a different opinion if the former moves towards the source of influence (DeGroot, 1974; French Jr, 1956). However, the literature on social psychology stipulates that if opinions are too distant, then the positive influence may not be accepted. Hence, the concept of bounded confidence has been introduced whereby only individuals espousing sufficiently close opinions may communicate (Deffuant et al., 2000). In turn, if individuals' opinions become more distant following communication, then such a mechanism is termed negative (aka dissimilative) influence (Macy et al., 2003). Note that in the case of a discrete opinion space, these assumptions are rather meaningless unless opinion values are ordered.

The positive influence mechanism explains elegantly how individuals reach agreement, and bounded confidence can model a situation when a social system is characterized by persistent disagreement (opinion fragmentation), whereas negative influence is one of the possible mechanisms explaining opinion polarization—the process in which individuals' opinions are stretched to the polar positions of the opinion space (Banisch & Olbrich, 2019). Thus, two camps of diametrically opposite opinions appear, a state of the social system that may have potentially dangerous consequences because it prevents democracy processes in general and consensus reaching in particular (Prasetya & Murata, 2020). An important challenge in the field of opinion dynamics models is to determine the settings under which the model would be able to generate stable opinion polarization (Flache et al., 2017). Possible solutions here, apart from negative influence, are mass media (Prasetya & Murata, 2020), social feedback processes (Banisch & Olbrich, 2019), social influence structure (Friedkin, 1999), arguments exchange (Banisch & Olbrich, 2021; Mäs & Flache, 2013), social identity (Törnberg et al., 2021), or mistrust (Adams et al., 2021).

## 3. Model

### *3.1. Model Setup*

We consider the system of $N$ agents that are connected by a social network $G$. Each agent's opinion may take one of $m$ values from the set $X = \{x_1, ..., x_m\}$ that represents a discrete opinion space, a construction that is extensively studied in the sociophysics literature (Axelrod, 1997; Castellano et al., 2009; Clifford & Sudbury, 1973; Galam, 1986). In some situations, one may assume that a binary relation is predetermined on the opinion space whereby types $x_1$ and $x_m$ stand for the most radical and polar position in that space:

$$x_1 < x_2 < \cdots < x_m.$$

Depending on the context, we will endow these quantities with different values.

In our model, the time is discrete; we denote the opinion of agent $i$ at time $t$ by $o_i(t) \in \{x_1, \ldots, x_m\}$. The population of agents having opinion $x_k$ at time $t$ is described by the quantity $Y_k(t) \in \{0, 1, \ldots, N\}$. Note that we assume that the system is "conservative" (agents do not leave the system, and there are no incoming agents): $\sum_{k=1}^{m} Y_k(t) = N$ for any $t$.

A key element of the model is a 3-D matrix $P = [p_{s,l,k}] \in \mathbb{R}^{m \times m \times m}$ where $s, l, k \in \{1, \ldots, m\}$, which governs opinion dynamics that unfolds on the social network. This matrix, which we will refer to as the *transition matrix* hereafter, prescribes probabilities of opinion shifts. Each opinion shift is a move in the opinion space that is a result of peer influence processes. The first index in $p_{s,l,k}$ stands for an agent's current opinion $x_s$, the second index describes the opinion $x_l$ of an influence source, and the last index represents the *potential* opinion $x_k$ of the target agent at the next time point. In other words, $p_{s,l,k}$ is the probability that an agent with opinion $x_s$ will switch their opinion to $x_k$ after being influenced by an agent holding opinion $x_l$:

$$p_{s,l,k} = \Pr[o_i(t+1) = x_k \mid o_i(t) = x_s, o_{i\leftarrow}(t) = x_l],$$

where $o_{i\leftarrow}$ denotes the opinion of the source of influence. As such, we should require

$$\sum_{k=1}^{m} p_{s,l,k} = 1$$

for any $s$ and $l$. Note that $p_{s,l,s}$ represents the probability of staying at the current position after interaction with opinion $x_l$. In the following, it will be convenient to represent different transition matrices by considering their slices over the first index. We will denote these slices, which are row-stochastic 2-D matrices, by $P_{s,:,:} \in \mathbb{R}^{m \times m}$. In brief, the matrix $P_{s,:,:}$ outlines the behavior of an agent who has opinion $x_s$. Its rows indicate the opinion of an influence source, and its columns represent potential opinion options:

$$P_{s,:,:} = \begin{bmatrix} p_{s,1,1} & \cdots & p_{s,1,m} \\ \cdots & \cdots & \cdots \\ p_{s,m,1} & \cdots & p_{s,m,m} \end{bmatrix}.$$

Thus, the number of parameters in the transition matrix depends only on the number of possible opinion values rather than on the total number of agents.

To illustrate the organization of the transition matrix, let us consider the following example.

***Example 1.*** *Consider the following transition matrix:*

$$P_{1,:,:} = \begin{bmatrix} 1 & 0 & 0 \\ 0 & 1 & 0 \\ 0 & 0 & 1 \end{bmatrix}, P_{2,:,:} = \begin{bmatrix} 0 & 1 & 0 \\ 0 & 1 & 0 \\ 0 & 1 & 0 \end{bmatrix}, P_{3,:,:} = \begin{bmatrix} 1/3 & 1/3 & 1/3 \\ 1/3 & 1/3 & 1/3 \\ 1/3 & 1/3 & 1/3 \end{bmatrix}.$$

*This transition matrix represents the likelihood of opinion shifts in opinion space $m = 3$. According to the transition matrix, an agent who has opinion $x_1$ is a conformist who completely follows the opinion of an influence source. In turn, agents who hold position $x_2$ are so-called stubborn agents who do not change their opinions in the presence of peer influence. Agents with opinion $x_3$ act in a purely random fashion regardless of who influences them.*

Now let us introduce how an opinion dynamics protocol is organized (see Figure 1). At each time point $t$, a randomly chosen agent $i$ is influenced by one of their neighbors $j$ in the social network (the neighbor is also chosen at random). Hence, the opinion of the focal agent $o_i(t)$ changes (or remains the same) according to the probability distribution established in the transition matrix. This influence mechanism is asymmetric: the opinion of agent $j$ does not change following the interaction.

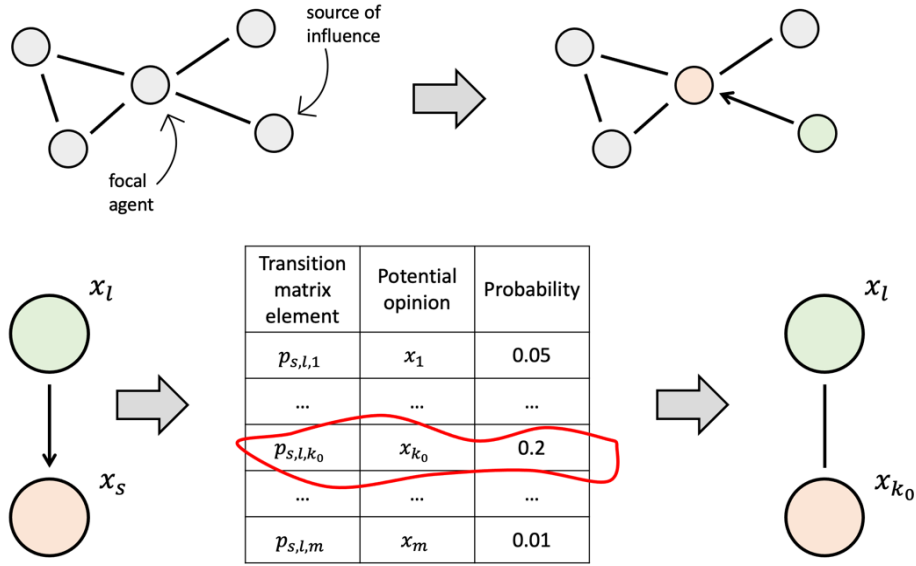

*Figure 1.* A graphical sketch of our model. The randomly chosen agent $i$ (orange node) having opinion $o_i(t) = x_l$ is influenced by one of their peers (green node) espousing opinion $o_j(t) = x_l$. The agent $i$'s new opinion is determined stochastically, according to the probability distribution $[p_{s,l,1}, \dots, p_{s,l,m}]$ established in the transition matrix.

*3.2. Flexibility of the model*

This model is extremely general and can capture a broad set of micro-influence mechanisms introduced in the literature. First, we demonstrate that the classic voter model (Clifford & Sudbury, 1973) can be easily obtained as a special case of our model.

**Example 2.** *Let us consider $m = 2$. To comply with established traditions whereby the binary opinion space is enriched with the spin interpretation, we denote $x_1 = -1$ and $x_2 = 1$. The following transition matrix*

$$P_{1,:} = P_{2,:} = \begin{bmatrix} 1 & 0 \\ 0 & 1 \end{bmatrix}$$

*establishes opinion dynamics that reproduces the voter model, in which randomly chosen agents alter their opinions to those of influence sources.*

By applying a binary relation to the opinion space (in other words, by introducing geometry in that space), we obtain an opportunity to encapsulate the concepts of bounded confidence and negative influence. We state that an agent makes a positive shift if their opinion moves towards the source of influence. A negative shift occurs if a user's opinion moves in the opposite direction.

***Example 3.*** *Let us assume that $m = 3$. Consider the following transition matrix slice:*

$$P_{2,:,:} = \begin{bmatrix} 0.75 & 0 & 0.25 \\ 0 & 1 & 0 \\ 0.25 & 0.25 & 0.5 \end{bmatrix}.$$

*In this case, the agent with the middle-side opinion $x_2$ may be influenced positively (with probability 0.75) and negatively (with probability 0.25) by opinion $x_1$. If influence comes from opinion $x_3$, then the focal agent may (i) hold their opinion (with probability 0.25), (ii) make a positive shift (with probability 0.5), and (iii) make a negative shift (with probability 0.25). In turn, if two agents with similar opinions $x_2$ communicate, then their opinions do not change.*

The notion of bounded confidence is based on the idea that an individual perceives (positive) influence only if the opinion of the influence source is not too far from their own opinion (Flache et al., 2017). Bounded confidence may take a strict form—that is, only agents with sufficiently similar opinions may influence each other—or a mild form, whereby agents with different opinions may communicate but with a small probability (Kurahashi-Nakamura et al., 2016; Mäs & Flache, 2013).

***Example 4.*** *Consider two slices of a transition matrix ($m = 4$):*

$$P_{1,:,:} = \begin{bmatrix} 1 & 0 & 0 & 0 \\ 0.5 & 0.5 & 0 & 0 \\ 1 & 0 & 0 & 0 \\ 1 & 0 & 0 & 0 \end{bmatrix}, P_{4,:,:} = \begin{bmatrix} 0 & 0 & 0.2 & 0.8 \\ 0 & 0 & 0.3 & 0.7 \\ 0 & 0 & 0.5 & 0.5 \\ 0 & 0 & 0 & 1 \end{bmatrix}.$$

*In this case, agents with opinion $x_1$ may be influenced by only those who have the nearest opinion $x_2$; they do not accept influence from more distant positions (strict bounded confidence assumption). Instead, agents who occupy the right edge of the opinion space may follow distant positions (by making a one-step opinion shift $x_4 \to x_3$) but with a decreasing rate (mild bounded confidence assumption).*

In Example 4, we do not pay attention to so-called *leapfrog* opinion shifts—situations when an agent's opinion $x_s$ moves towards the opinion of an influence source $x_l$, with a magnitude that is higher than the distance between $x_s$ and $x_l$. In this case, the focal agent's opinion skips the influence source's one. Leapfrog opinion shifts are rarely considered in the theoretical studies but nonetheless may be encountered in empirical environments (Friedkin et al., 2021; Kozitsin, 2020, 2021). In principle, such situations may be attributed to measurement errors (Carpentras & Quayle, 2021).

Acting in a similar fashion, one could adjust the values of the transition matrix to represent more complex microscopic assumptions on social influence, such as *moderated* positive influence or combinations of positive and negative influence in which coexistence may take quite nontrivial forms (Kozitsin, 2021; Takács et al., 2016).

If the binary relation on the opinion space is introduced, then the limit $m \to \infty$ provides an approximation of a continuous opinion space (without loss of generality, we may consider the interval $[0,1]$), which has gained substantial attention in the literature (Mastroeni et al., 2019).

The flexibility of our model is not without limitations. One can notice that the model assumes that agents with similar opinions should act equally on average in similar situations (that is, being exposed to comparable influence opinions), an assumption that significantly reduces the model's predictive power because not all ties transmit influence on an equal basis. In other words, our model can reproduce only the average patterns of opinion formation processes and approximate only anonymized forms of continuous opinion models (whereby all influence weights are equal). This issue makes the model less flexible than, for example, the DeGroot model, which allows individuals to allocate different influence weights to their peers.

On the other hand, our model can easily explain the situation when an agent acts differently (by choosing positive or negative shift) as a response to the same influence opinion. This ability may be (i) due to the stochasticity of the model and (ii) because the model allows encoding of different opinion-changing strategies, depending on the current opinion of the focal individual. In the case of the DeGroot model (which is linear in its canonic form), the same can be achieved if one introduces nonlinearity in the influence matrix, an assumption that makes the model significantly more sophisticated.

*3.3. Model identification*

To calibrate the elaborated model, one needs to know (1) the trajectories of individual opinions and (2) the history of the individuals' communications. To be more precise, for a given individual $i$, one must know their opinion $y_i$ before communication, the opinion of the influence source $y_{i\leftarrow}$, and the focal agent's opinion $z_i$ after communication. One should first apply the procedure of discretization on experimental opinions if these opinions are initially continuous. Here, one should find the most appropriate discretization step, which should be a sort of compromise: a step that is too large could lead to losing potentially useful information on individuals' opinion trajectories, whereas too small a step results in a sharp increase in the number of transition matrix elements, which are now highly difficult to interpret and gives way to unnecessary data fluctuations. The resulting discrete opinions (for convenience, we denote them similarly) can be used to estimate the transition matrix elements:

$$p_{s,l,k} = \frac{\#\{i \mid (y_i = x_s) \& (y_{i\leftarrow} = x_l) \& (z_i = x_k)\}}{\#\{i \in I \mid (y_i = x_s) \& (y_{i\leftarrow} = x_l)\}},$$

where $\#\{...\}$ denotes the cardinal number of the set. To put it simply, $p_{s,l,k}$ is computed as the fraction of individuals who made opinion change $x_s \to x_k$ among those whose opinion is $x_s$ and are influenced by opinion $x_l$. To compute all the transition matrix components, one needs to be provided with a substantial opinion diversity, which ensures that all combinations of $x_s$ and $x_l$ are represented in the data. Otherwise, the available statistics would be insufficient to calibrate the transition matrix. Individuals' opinions should be represented at least twice in the data: before and after interactions. However, longer opinion trajectories will be useful because they provide more room for analysis.

After the transition matrix is estimated, one should first analyze it to understand the average patterns of opinion dynamics and then make predictions about the future behavior of the system. Examples of the transition matrices estimated from empirical data will be presented below in this paper.

Note that we do not impose any requirements on the nature of empirical opinions. They could be discrete—in this case, we will consider low-dimensional transition matrices, or continuous—then, we firstly discretize opinions. In principle, the history of an individual's communications (with whom they talked), which is highly difficult to retrieve in non-laboratory settings, can be replaced by more simple forms of structures of social connections, such as a friendship network, which could be relatively easily retrieved from the Web. Of course, more detailed information on how individuals interact with each other will make the model more precise, but in the following sections, we will demonstrate that even a simple friendship network may serve as a good approximation of the actual communication network in the sense that it could simulate artificial social systems consistent with empirics at the macro scale.

## 4. Analytical results
### *4.1. Mean-field approximation*

The model elaborated above can be studied using mean-field approximation. Within this section, we assume that the social network is a complete graph whereby each agent can communicate with each one.

Establishing scaled time $\tau = \frac{t}{N}$ and scaled time step $\delta\tau = \frac{1}{N}$, for large $N$ we get the nonlinear autonomous system of differential equations (see Appendix for details):

$$\frac{dy_f(\tau)}{d\tau} = \sum_{s,l,k} y_s(\tau) y_l(\tau) p_{s,l,k}(\delta_{k,f} - \delta_{s,f}), \quad f \in \{1, \ldots, m\}, \qquad (1)$$

which should be equipped with the initial condition:

$$y_f(0) = y^f, \quad f \in \{1, \ldots, m\}, \qquad (2)$$

where $\sum_{f=1}^{m} y^f = 1$ and $y^f \in [0, 1]$. Note that one of the equations in (1) is redundant.

Equilibrium points of the system (1) are given by:

$$\begin{cases} g_f = 0, & f \in \{1, \ldots, m\}, \\ y_1 + \cdots + y_m = 1, \\ y_f \in [0,1], & f \in \{1, \ldots, m\}, \end{cases}$$

where we use the notation:

$$g_f = \sum_{s,l,k} y_s y_l p_{s,l,k}(\delta_{k,f} - \delta_{s,f}).$$

Due to the fact that the right side of (1) is a polynomial, we can guarantee that the Cauchy problem (1)^(2) has a unique solution, which is an analytic function of parameters $p_{s,l,k}$ and $y^f$. In other words, small changes in $p_{s,l,k}$ and $y^f$ lead to small perturbations of the solution.

In the following two subsections, we will exemplify the elaborated analytical results with the low-dimension cases $m = 2$ and $m = 3$, enriching them with data describing real opinion dynamics processes.

*4.2. Analysis of the binary opinion space*

If $m = 2$, which corresponds to the binary opinion space, then system (1) takes the quite simple form:

$$\begin{cases} \dot{y}_1 = -p_{1,1,2}y_1^2 + (p_{2,1,1} - p_{1,2,2})y_1y_2 + p_{2,2,1}y_2^2, \\ \dot{y}_2 = p_{1,1,2}y_1^2 + (p_{1,2,2} - p_{2,1,1})y_1y_2 - p_{2,2,1}y_2^2. \end{cases} \quad (3)$$

One of two equations in system (3) is unnecessary. Substituting $y_1 = 1 - y_2$ into the first one, we get:

$$g_1 = ay_1^2 + by_1 + c$$

where:

$$a = -p_{1,1,2} + p_{1,2,2} - p_{2,1,1} + p_{2,2,1},$$
$$b = p_{2,1,1} - p_{1,2,2} - 2p_{2,2,1},$$
$$c = p_{2,2,1}.$$

If $y_1^* \in [0,1]$ is an equilibrium point, then it should admit equation:

$$g_1(y_1^*) = 0.$$

The sign of the quantity:

$$\dot{g}_1(y_1^*) = 2ay_1^* + b$$

determines the stability properties of this equilibrium.

Let us now equip the model with empirical data from the dataset that was recently reported in Kozitsin (2020, 2021), where the author analyzed longitudinal data representing the dynamics of (continuous) opinions of a large-scale sample of OSN users (hereafter – Dataset). Using the first two opinion snapshots[1] from Dataset (see Appendix for details), we obtain the following transition matrix:

$$P_{1,:,:} = \begin{bmatrix} 0.975 & 0.025 \\ 0.952 & 0.048 \end{bmatrix}, P_{2,:,:} = \begin{bmatrix} 0.066 & 0.934 \\ 0.049 & 0.951 \end{bmatrix}. \quad (4)$$

One can observe that transition matrix (4) is very different from the one that represents the voter model (see Example 2); in the current case, agents rarely change their opinions if they are exposed to challenging positions. In turn, they have a nonzero chance of their opinion changing after being exposed to the same position—a phenomenon that was referred to in (Krueger et al., 2017) as *anticonformity*. Nonetheless, the likelihood of an opinion shift slightly increases if two agents with opposite positions communicate, compared to when they have similar opinions.

---

[1] Note that Dataset includes three opinion snapshots overall. All results presented below remain virtually the same if one uses different snapshot sequences (e.g., first and third or second and third).

System (3) has only one meaningful (located in the unit square) equilibrium point $y_1^* \approx 0.644, y_2^* \approx 0.356$, which is asymptotically stable because $g'(y_1^*) < 0$.

### 4.3. Analysis of the triple opinion space

Let us now increment the dimensionality of the opinion space by one and consider triple opinion space $X = \{x_1, x_2, x_3\}$. In this configuration, opinions $x_1$ and $x_3$ stand for antagonistic positions, whereas $x_2$ is somewhat neutral, located in the center. For brevity, we do not describe how system (1) appears in this case. Instead, we immediately obtain th0e transition matrix from empirics in a similar manner to the process performed in the previous subsection:

$$P_{1,:,:} = \begin{bmatrix} 0.96 & 0.04 & 0 \\ 0.942 & 0.057 & 0.001 \\ 0.907 & 0.091 & 0.002 \end{bmatrix}, P_{2,:,:} = \begin{bmatrix} 0.039 & 0.952 & 0.008 \\ 0.021 & 0.969 & 0.01 \\ 0.02 & 0.944 & 0.036 \end{bmatrix}, P_{3,:,:}$$

$$= \begin{bmatrix} 0.001 & 0.082 & 0.916 \\ 0.001 & 0.07 & 0.929 \\ 0.001 & 0.054 & 0.945 \end{bmatrix}. \qquad (5)$$

The slices of the transition matrix presented above reflect several remarkable features. First, larger values of opinion difference between individuals increase the rate of positive influence. One can observe this trend by inspecting slices $P_{1,:,:}$ and $P_{3,:,:}$: for example, the second and third columns in matrix $P_{1,:,:}$ are established so that their values increase as the index of the columns also rises. Further, individuals located in the middle of the opinion space may make negative shifts: $p_{2,1,3} > 0, p_{2,3,1} > 0$. However, according to (5), positive shifts are more likely than negative ones: $p_{2,1,3} < p_{2,1,1}, p_{2,3,1} < p_{2,3,3}$.

System (1) augmented with transition probabilities (5) has only one meaningful (located in the unit square) equilibrium point that can be obtained graphically or via solving the following optimization problem (see Figure 2):

$$\begin{cases} \left(g_1(y_1, y_2)\right)^2 + \left(g_2(y_1, y_2)\right)^2 \to \min_{y_1, y_2}, \\ y_1, y_2 \in [0,1], \\ y_1 + y_2 \leq 1. \end{cases} \qquad (6)$$

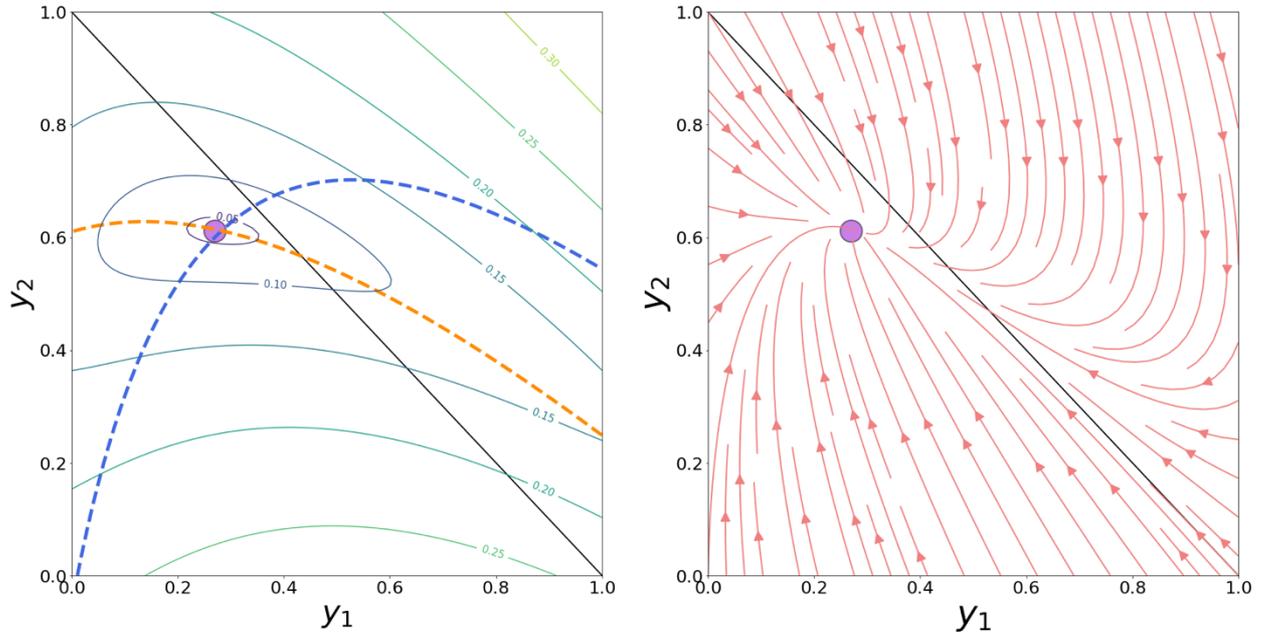

*Figure 2.* (Left panel). In this panel, we plot contour lines of the objective function in (6) (raised to the power of 1/4 for demonstration purposes). The violet circle represents the minima obtained via solving optimization problem (6). Instead, dashed lines portray the curves $g_1(y_1, y_2) = 0$ (blue) and $g_2(y_1, y_2) = 0$ (orange). Their intersection marks the equilibrium point, which coincides with the numerical solution of the optimization problem. The black solid line plots $y_1 + y_2 = 1$. We are interested in the area $y_1 + y_2 \leq 1$ beneath this line. (Right panel). The phase portrait of the system demonstrates that it has a unique attractor (asymptotically stable equilibrium point). The behavior of phase curves near the point is well approximated by the Jacobian matrix.

We find that $y_1^* \approx 0.269$, $y_2^* \approx 0.611$ (and $y_3^* \approx 0.12$ correspondingly). The Jacobian matrix at the equilibrium point has two different negative eigenvalues; hence, we can identify the behavior of the phase curves near the equilibrium point, which is an asymptotically stable nodal sink.

**5. Numerical experiments**

We perform extensive numerical experiments to investigate the behavior of our model and, more specifically, support the analytical results derived in Section 4. Besides, we would like to understand whether our model can generate artificial social systems close (in some respects) to those observed empirically. We recognize that it is unlikely that we will be able to predict the opinion trajectories of specific individuals. However, we hope to predict the system's dynamics at the macroscopic level.

We build our further analysis on the investigation into how the model works on synthetic random networks by employing the following macroscopic metrics.

**M1**. The fraction of individuals $y_i$ who have opinion $x_s$ for $s \in \{1, ..., m\}$. The combination of variables $y_1(t), ..., y_m(t)$ represents public opinion at time $t$. In what follows, we will refer to them as public opinion variables.

**M2**. *Assortativity coefficient*

$$C(A, o(t)) = \frac{\sum_{i,j} \left(a_{ij} - \frac{k_i k_j}{2q}\right) o_i(t) o_j(t)}{\sum_{i,j} \left(k_i \delta_{ij} - \frac{k_i k_j}{2q}\right) o_i(t) o_j(t)} \in [-1,1], \qquad (7)$$

which measures whether the system at hand is homophilic (connected nodes tend to have similar opinions). In (7), vector $o(t) = [o_1(t) \ ... \ o_N(t)]^T$ stands for current agents' opinions, adjacency matrix $A = [a_{ij}] \in \{0,1\}^{N \times N}$ describes the stkructure of social network $G$, and $q$ is the number of edges in the network. $k_i$ represents node $i$'s degree: $k_i = \sum_{i=1}^{N} a_{ij}$. To put it simply, (7) measures how similar the neighboring opinions are, compared with the configuration in which edges are placed at random (M. E. J. Newman, 2003). For homophilic networks (most empirically observed social networks are homophilic), metric (7) takes positive values (assortative mixing).

**M3**. The *dissimilarity coefficient* $D(o(t))$, which effectively assesses the current level of polarization (Banisch & Olbrich, 2019; Flache & Macy, 2011). This measure is defined as the standard deviation of all pairwise opinion distances and takes values in the range between $D = 0$ (no polarization – all opinions are equal) and $D = 1$ (the highest level of polarization – individuals are divided into two equally sized camps located on the edges of the opinion space), provided that opinions lie in the interval $[-1, 1]$.

## *5.1. Simulation settings*

In experiments, we consider $N = 2000$ agents, who are endowed with randomly generated initial opinions. Unless otherwise stated, opinions are initialized from the generalized Bernoulli distribution in which we set the uniform vector of probabilities ($y_i(0) = 1/m$ for $i \in \{1, \dots, m\})^2$. For large $m$, the initial opinion configuration is characterized by $C \approx 0, D \approx 0.5$. Note that in the case of the complete graph, the assortativity coefficient is always equal to null. In each experiment, a new network is generated, as well as new initial opinion values. Apart from the complete graph model (under which we derived the mean-field approximation and which we use for the public opinion dynamics analysis only), we employ four synthetic graph models that are widely used in the social simulations literature (Giardini & Vilone, 2021; Perra & Rocha, 2019): (i) Erdős–Rényi network, (ii) random geometric network, (iii) Watts–Strogatz network, and (iv) Barabási–Albert network. Detailed information on network configurations is presented in Table 1. Experiments typically lasted no more than one million iterations, a time interval that is sufficiently large to inspect the model's behavior. We repeated each experiment 20 times to gain more precise estimations.

Table 1

---

[2] We tested different initial opinion configurations; however, we found that they have no influence on the asymptotic behavior of the model.

Properties of synthetic networks

| Network model | Parameter(s) | Brief description |
|---|---|---|
| Erdős–Rényi | 15000 edges | The model places a predefined number of edges between the nodes at random. The resulting network has no common features with real social graphs (ignoring sparsity that can be obtained for small values of the parameter) but may serve as their simplest (apart from the complete graph) approximation. |
| Random geometric graph | Threshold distance 0.05 | The nodes are randomly placed in the unit square. Each pair of nodes is connected if and only if they are distant for no more than the threshold value. The model allows us to obtain the "structured" graphs containing communities (organized geographically), the presence of which is a prominent signature of real social networks. |
| Watts–Strogatz | Each node is connected to 15 nearest ones, probability of rewiring $p_{rew} = 0$ (WS1) | Initially, nodes are placed in the ring topology, in which each node is connected to a predefined number of nearest ones. Edges are rewired at random with the probability $p_{rew}$. If $p_{rew} = 0$, then we obtain clustered networks (with an average clustering coefficient $\approx 0.69$) that feature high values of average path length ($\approx 72$). A tiny increase in the value of the rewiring probability ($p_{rew} = 0.01$) leads to networks that are still highly clustered (average clustering $\approx 0.67$) but characterized by a small average path length ($\approx 7.6$) – so-called small-world networks. If $p_{rew} = 1$, then the model generates graphs with no clustering but with even lower values of the average path length ($\approx 3$). |
| | Each node is connected to 15 nearest ones, $p_{rew} = 0.01$ (WS2) | |
| | Each node is connected to 15 nearest ones, $p_{rew} = 1$ (WS3) | |

| | | The model adds new nodes in the system |
| :---: | :---: | :--- |
| Barabási–Albert | Each new node is attached to seven already existing ones | sequentially, connecting them with existing ones at random, following the preferential attachment rule. The resulting networks follow the power law degree distribution, which is widely observed in real social networks. |

*Note*: all model parameters are tuned to ensure that the resulting networks are connected and have approximately the same density (or the same number of edges $q$).

## 5.2. Transition matrices

We estimate the transition matrix using information from Dataset and concentrate on the first two opinion snapshots (see Appendix for details). We analyze cases $m = 2$, $m = 3$, and $m = 10$. The first two opinion space configurations require only a small number of variables to be parametrized, and this ability is useful in interpretations and for demonstrative purposes. Instead, the tenfold opinion space provides a more precise approximation of the underlying social processes. Further increase in $m$ may lead to unnecessary fluctuations in the data and a sharp increase in the number of transition matric elements. Transition matrices for binary and triple opinion spaces have already been introduced in (4) and (5). The tenfold transition matrix is partially presented (and discussed) in the Appendix; its full representation can be found in the Online Supplementary Materials.

## 5.3. Measurements, hypotheses, and expectations

The immediate observation that could be made from the estimated transition matrices is that the system has no stable states at the microscopic level: for every opinion vector, there is a nonzero probability that a randomly chosen agent will change their opinion at the next time point (even after exposure to the same opinion). However, in the following, we will demonstrate that the system has a stable converging tendency from the perspective of the macroscopic metrics.

Kozitsin (2020, 2021) reported that the social system under consideration is homophilic with an assortativity coefficient of approximately 0.14, which may be considered as a not particularly strong (bot noticeable) rate of homophily. He also observed that users' opinions tend to stretch out to the edges of the opinion space in such a way that the fraction of individuals having middle-located (moderate) opinions decreases, individuals disposed on the left edge grow in number, and right-opinion users tend to keep their number or slightly decrease. To gain a more systematic understanding of the system, we calculate the metrics **M1**–**M3** using different discretization strategies for all three opinion snapshots from Dataset (see Table 2). We observe the following dynamical patterns: (i) growth of the population of individuals espousing left-side opinions, decrease of those who hold a middle-side opinion, and a relatively small decrease of right-side opinion persons; (ii) extremely small increasing trend in the homophily level; (iii) extremely small increasing trend in the polarization rate. As such, we hypothesize that the model calibrated on the same data should be able to achieve similar metric values (hereafter – reference values) at some point of its evolution and, further, demonstrate the same dynamics patterns near this point.

Table 2

Values of metrics **M1**–**M3** drawn from Dataset

| Metric | Discretization step | | | | | | | | | |
|---|---|---|---|---|---|---|---|---|---|---|
| | $m = 2$ | | | | $m = 3$ | | | | | $m = 10$ |
| Public opinion variables | $y_1$ | 0.568 | 0.576 | 0.582 | $y_1$ | 0.161 | 0.168 | 0.174 | | |
| | $y_2$ | 0.432 | 0.424 | 0.418 | $y_2$ | 0.702 | 0.697 | 0.692 | | |
| | | | | | $y_3$ | 0.137 | 0.135 | 0.134 | | |

| | | | | | | | | | |
|---|---|---|---|---|---|---|---|---|---|
| Assortativity coefficient | 0.109 | 0.109 | 0.109 | 0.106 | 0.107 | 0.108 | 0.141 | 0.141 | 0.141 |
| Dissimilarity coefficient | | | | 0.575 | 0.585 | 0.585 | 0.356 | 0.366 | 0.357 |

*Note:* the dissimilarity coefficient in the case of the binary opinion space is not useful and, therefore, we do not calculate it. The dynamics of public opinion variables in the case of the tenfold opinion space are too massive—one can find this information in Kozitsin (2021) if necessary.

## 6. Results

### *6.1. Macroscopic behavior of the model*

Our experiments reveal that regardless of the network topology, the behavior of public opinion variables remains the same. The evolution of the model can be decomposed into two periods (see Figure 3, panels A, B). In the first one, populations of camps $y_1(t), \ldots, y_m(t)$ nearly monotonically converge to the theoretical predictions $y_1^*, \ldots, y_m^*$ (see Subsections 4.2, 4.3). In the next period, the system fluctuates around these limiting values.

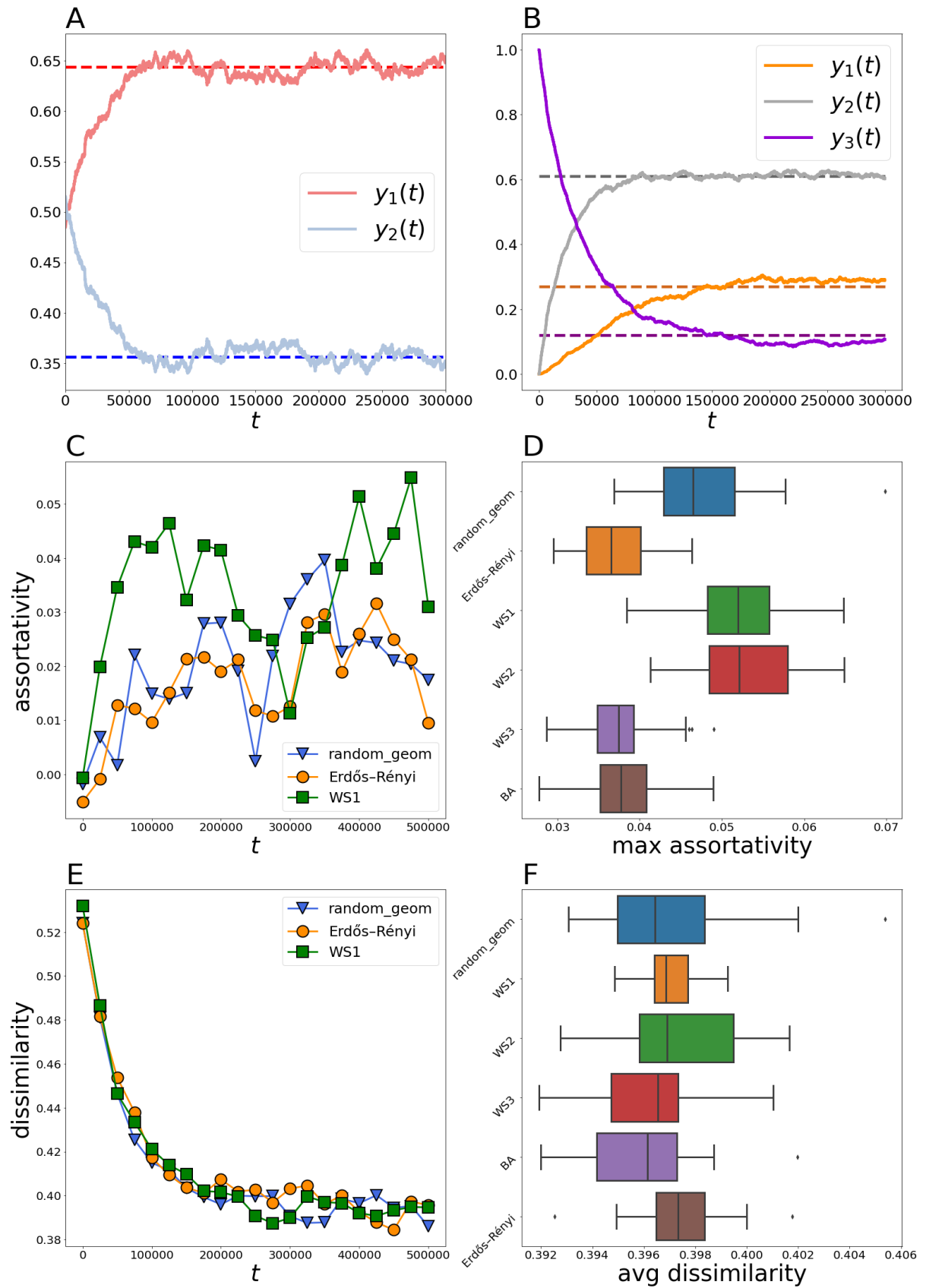

*Figure 3*. Ideal-typical evolution of public opinion variables in the model in the case of binary (panel A, initial opinions are drawn from the uniform distribution) and triple (panel B, initial opinions are equal to $x_3$) opinion spaces. Dashed lines plot theoretically predicted

equilibrium values. (Panel C). Ideal-typical behavior of the assortativity coefficient in experiments ($m = 10$). The network topology with no clustering (Erdős–Rényi) exhibits lower assortativity values than clustered ones (random geometric and WS1). (Panel D). The panel plots how the maximal value of assortativity during a single experiment varies with the network topology ($m = 10$). (Panel E). Ideal-typical dynamics of the polarization coefficient for different network topologies ($m = 10$). (Panel F). The average (over time) value of the dissimilarity coefficient after the system has been stabilized ($t \geq 250{,}000$) during a single experiment as a function of the network topology ($m = 10$).

At the beginning, the system is characterized by the almost zero assortativity because opinions are endowed at random. After a simulation has been initiated, the system rapidly becomes homophilic and then features fluctuations in a positive area demonstrating a relatively low rate of homophily (see Figure 3, panel C). For example, for the binary opinion space, the maximal assortativity rate observed is 0.05 (under WS1 topology), a value that is far from the empirical reference value (0.109). A similar difference in assortativity values was discerned in the case of the tenfold opinion space (0.07 in simulations against reference value 0.141). Further, we found that more clustered networks (random geometric, WS1, and WS2) tend to produce more homophilic systems (see Figure 3, panel D).

The typical behavior of the polarization coefficient can be easily predicted because we know how public opinion evolves: starting from some point (that is determined by the initial opinion distribution), the polarization coefficient should firstly drift to the limiting value that characterizes the polarization of the stationary state opinion distribution $\{y_1^*, \ldots, y_m^*\}$. Depending on the initial opinion configuration, this stage of evolution may feature growth (if the initial polarization level is lower than the limiting value) or decrease (if the initial polarization rate exceeds the limiting one). After the limiting value is achieved, the polarization coefficient should fluctuate around it. Further, this limiting value should not depend on the network topology because the latter does not affect the stationary state opinion distribution. Our numerical experiments (see Figure 3, panels E, F) confirm this proposal. We do not observe any relation between the network topology and the system's asymptotic polarization.

The presented results indicate that the elaborated model demonstrates a stable converging tendency: its macroscopic parameters tend to some limiting values at first and then feature oscillations around them. Depending on a particular macroscopic metric, corresponding limiting values may (the assortativity coefficient) or may not (public opinion variables, dissimilarity) be affected by the network topology. However, all of them are not sensitive to the initial opinion configuration. These findings contradict the work of Banisch and Olbrich (2019) and Stern and Livan (2021), who reported that the community organization of the network is one of the key factors of polarization. Instead, we found that this organization makes the network more homophilic. In the limit $t \to \infty$, the system features opinion fragmentation, which is characterized by persistent disagreement between agents. Further, one can also notice opinion polarization if the system begins from a densely concentrated opinion distribution. For example, if opinions are concentrated near the center of the opinion space, in this case, at the initial stage of the system's evolution, opinions will be prone to antagonistic or anticonformity-based interactions (in Kozitsin (2020), this phenomenon was attributed to the striving for uniqueness) and, thus, will move towards the extreme opinion values $x_1$ and $x_m$. However, increasing pairwise distances between agents' opinions will give way to assimilative interactions that are less likely to occur if opinions are too close. This process will continue until a sort of balance between assimilative and antagonistic interactions is reached.

The model exhibits good predictability from the perspective of the public opinion dynamics: the mean-field approximation derived in Section 4 is plausible not only for complete graphs (under the assumption of which this approximation was obtained) but also in other settings whereby social connections may have a quite complex structure. Next, the model can reproduce positive assortativity (i.e., can create a homophilic social system). However, the observed homophily rates are far lower than the empirical reference values. Further, dissimilarity demonstrated by the system after stabilization (which is slightly more than the corresponding reference values) provides a clue that the system can reach the reference polarization level if its initial degree of polarization is lower than the asymptotic one—this situation could arise if initial opinions are densely concentrated (for example, near the center of the opinion space).

To understand whether the simulation system can demonstrate similar behavior to the empirically observed one (see Table 2), we use the triple opinion space because it requires only a few macroscopic metric values to be analyzed (for example, in the case $m = 10$, we need to inspect the behavior of 12 variables). Quite close matching between simulations and empirics can be achieved if one begins a simulation run from opinion distribution $y_1(0) = 0.07, y_2(0) = 0.79, y_3(0) = 0.14$ (see Figure 4). In this case, at moment $t \approx 20{,}000$, we notice that metrics **M1** and **M3** match the reference values as well as their local dynamic patterns (see Table 2)[3]. An exception is the assortativity coefficient, the simulation values of which are below the reference one. These findings lead us to the following question:

*What modifications should one add to the model to make it possible to reproduce empirically observed homophily?*

---

[3] It is not particularly surprising that the model can create fragmented and polarizing social systems because the corresponding transition matrix (5) allows negative interactions between agents.

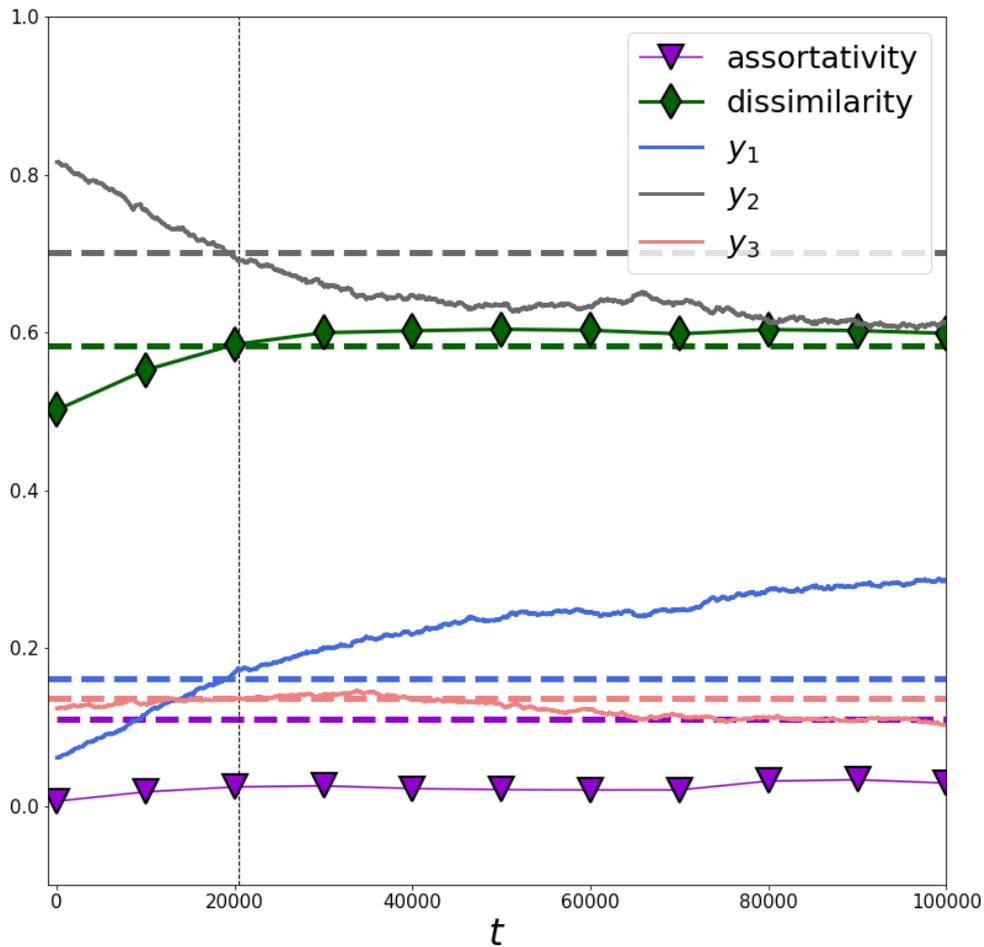

*Figure 4.* Ideal-typical behavior of the system in the case of the triple opinion space if initial opinions are drawn from the distribution $y_1(0) = 0.07, y_2(0) = 0.79, y_3(0) = 0.14$. Dashed horizontal lines plot reference values from Table 2 (we take values from the first opinion snapshot). Common colors indicate the same metrics. The vertical dashed line marks the point of time at which all macroscopic metrics except **M2** tend to coincide with the empirical reference values. One can observe a sharp discord in assortativity values between empirics (violet dashed line) and simulations (violet curve with triangle markers).

## *6.2. Possible modifications*

Let us present some possible explanations of the observed discord in the values of the assortativity coefficient between simulations and empirics and feasible avenues for resolving this conflict.

**Methodological errors in Dataset.** A discerned divergence between numerical experiments and empirics may stem from methodological errors in obtaining the underlying empirical data. Because these data were derived through a natural experiment in which individuals' opinions were estimated by using some heuristics (more precisely, it was assumed that users' opinions are reflected by the information sources they are subscribed to), it could mean that the empirical reference values we strive to equal are incorrect. Unfortunately, we cannot fix this problem; hence, we leave this situation beyond the scope of this article and assume that the reference values are identified correctly.

**Noise in the estimated transition matrices.** A slightly different idea is to suppose that the reference values of macroscopic metrics *are correct* (i.e., individuals' opinions were estimated faithfully), but the transition matrix is identified *with errors*. The point is that errors in opinion identification naturally lead to mistakes in the estimated transition matrix (this is precisely what the previous paragraph discusses); however, to identify the transition matrix accurately, apart from the opinion values knowledge, we should also be able to determine the influence individuals are exposed to (see Subsection 3.3. and Appendix for details). This problem requires uncovering the influence network (Ravazzi et al., 2021), which is also a challenging task. Note that in Dataset, all influence weights are assumed to be equal (the influence opinion directed on a user is the average of opinions of the user's friends), an assumption which is unlikely true: some ties may be more successful in social influence transmission (Bond et al., 2012). Further, an influence system retrieved from OSN is likely incomplete because it neglects to consider the influence beyond the online world (more precisely, beyond a particular OSN). One more effect stems from our algorithm of the transition matrix identification and the nature of data (see Appendix for details): opinion dynamics presented in Dataset are identified under the assumption of many-to-one interactions, whereas the model assumes one-to-one interactions. Besides, such factors as selectivity or personalization algorithms (see below) may dominate. As a result, the real transition matrix may differ from the estimated one.

To model mistakes in the estimated transition matrix, we consider the binary opinion space, in which these mistakes may be parametrized by two variables:

$$P_{1,:,:} = \begin{bmatrix} 0.975 + \alpha & 0.025 - \alpha \\ 0.952 - \beta & 0.048 + \beta \end{bmatrix}, P_{2,:,:} = \begin{bmatrix} 0.066 + \beta & 0.934 - \beta \\ 0.049 - \alpha & 0.951 + \alpha \end{bmatrix}. \quad (8)$$

In (8), variables $\alpha$ and $\beta$ represent small perturbations that stand for the difference between the estimated transition matrix and the real one. In the general case, we should use four different variables to parametrize noise. However, due to technical reasons, we have decided to consider "symmetric" disturbances. Using this approach, we analyze how the values of $\alpha$ and $\beta$ affect the system's assortativity. We investigate domain $\{-0.02 \leq \alpha \leq 0.02, -0.02 \leq \beta \leq 0.02\}$ with step 0.005 across both dimensions. Assuming $\alpha = 0$ and $\beta = 0$, we return to the earlier-obtained transition matrix (4).

**Selectivity** is a well-documented tendency of social actors that we could not ignore, and it creates ties with those having similar opinions and breaks connections that promote uncomfortable information (Holme & Newman, 2006; Lewis et al., 2012; Neubaum et al., 2021; Sasahara et al., 2021). Along with (assimilative) social influence, selectivity is considered to be a main driver that makes social networks homophilic. As such, we may hypothesize that by adding selectivity into the model, we will increase the level of homophily, which is one of our purposes. However, the question is how selectivity affects polarization.

There is a line of research devoted to modeling coevolutionary processes whereby social influence mechanisms are combined with the dynamics of social graphs driven by selectivity (Frasca et al., 2019; Holme & Newman, 2006). We incorporate selectivity into our model by introducing parameter $\gamma \in [0,1]$ (*selectivity rate*), which operates as follows. At each time point $\tau$, we select an agent $i$ (with opinion $o_i(t)$) and one of their neighbors $j$ (having opinion $o_j(t)$) at random. If their opinions are not too distant ($|o_i(t) - o_j(t)| \leq \Delta o$), then they follow the standard opinion dynamics protocol: agent $i$ changes their opinion in accordance with the transition matrix. Otherwise (if $|o_i(t) - o_j(t)| > \Delta o$), with probability $\gamma$ tie $(i, j)$ is deleted and a new tie appears: agent $i$ creates a connection with a random (not neighboring) vertex $k$ whose opinion lies within interval $[o_i(t) - \Delta o, o_i(t) + \Delta o]$. If there are no such vertices[4], then tie $(i, j)$ does not disappear (nothing happens). In the contrary case (with probability $1 - \gamma$), the standard opinion dynamics protocol is implemented. Thus, at each iteration, either an opinion changes, or the network evolves, or nothing happens. Note that the number of ties in the evolving social network remains the same.

**Personalization systems** are a signature of the contemporary social networking services. Due to our inability to capture all the information from these sites that we would like to obtain, personalization algorithms sort the online content and suggest to us those that will be most important (to us), from the algorithms' point of view, of course. Personalization algorithms may significantly influence the macroscopic behavior of opinion dynamics processes because they control information flows between individuals. Relatively recently, scholars began to combine ideas of personalization algorithms with opinion formation models (De Marzo et al., 2020; Maes & Bischofberger, 2015; Perra & Rocha, 2019; Rossi et al., 2019). Their results indicate that personalization may both amplify and reduce polarization, subject to the underlying opinion formation model. Further, scholars argue that personalization algorithms may amplify the formation of echo chambers and, thus, increase the level of homophily. Besides, it is unlikely that the opinion dynamics presented in Dataset were not affected by personalization algorithms. On this basis, we need to incorporate personalization in our model. For our purposes, we employ one of the simplest approaches whereby communications between individuals may be declined. More precisely, selected agents $i$ and $j$ do not communicate (and the system goes to the next time step) with probability $\delta$ (*personalization rate*) if their opinions differ for more than $\Delta o$.

---

[4] It may be, for example, if the focal agent is the last representative of the opinion camp.

**Implementation details.** We implement assortativity and personalization into the model only in the cases of the triple and tenfold opinion spaces. Importantly, we combine them: if selectivity and personalization rates are positive at once, then agents can both change their connections and be affected by the personalization algorithm, as occurs in the following fashion (see Figure 5). First, the personalization algorithm checks whether $|o_i(\tau) - o_i(\tau)|$ is greater than $\Delta o$ or not. If the latter is true, then communication between agents is allowed, and $i$ changes their opinion as usual. Otherwise (if opinions are too distant), the personalization algorithm activates and prohibits the communication with probability $\delta$, and nothing happens (the system goes to the next time step). In contrast, with probability $1 - \delta$ communication between agents $i$ and $j$ is allowed. This permitted communication can go in two different ways. In the first one (that occurs with probability $\gamma$) agent $i$ decides to renew their social environment by replacing tie $(i,j)$ because it makes agent $i$ uncomfortable. The second direction implies that agent $i$ accepts influence from agent $j$ and follows the standard opinion dynamics protocol (whatever it leads to). Note that if we set $\gamma = 0$ and $\delta = 0$ in the resulting model (Model 2), we return to the previously elaborated model (Model 1).

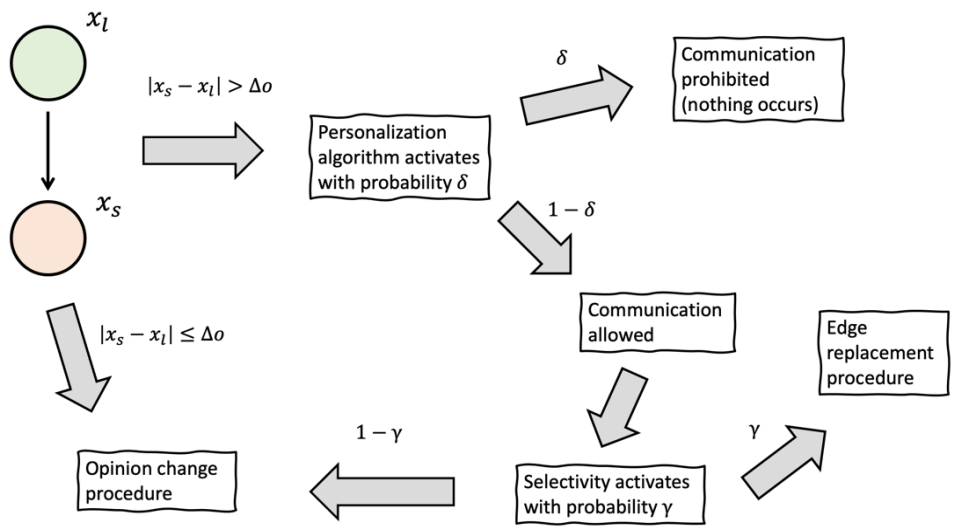

Figure 5. A graphical sketch of Model 2. The interaction of agents $i$ and $j$ with opinions $x_l$ (the influence source) and $x_s$ (influenced agent) respectively is driven by personalization

(that may exclude the communication) and selectivity (that may lead to the replacement of edge $(i, j)$ instead of the standard procedure of opinion change).

We employ the idea of noise in the transition matrix only for the binary opinion space because it requires (under the symmetrical assumption on noise) only two parameters to be used. Hence, we do not investigate how the data noisiness affects polarization patterns and concentrate only on the behavior of the assortativity coefficient.

Model 2 is investigated only under the empirically calibrated transition matrix derived for triple and tenfold opinion spaces. This approach gives us the opportunity to analyze the effects of selectivity and personalization factors on the dynamics of homophily and polarization. In Model 2, we use threshold $\Delta o = 0$ if $m = 3$ (which is applied to opinion values $x_1 = 0, x_2 = 1, x_3 = 2$) and threshold $\Delta o = 3$ if $m = 10$ (which is applied to opinion values $x_1 = 0, \ldots, x_{10} = 9$).

We recognize that the implementation of selectivity and personalization affects not only the opinion dynamics itself but also influences the transition matrix that we observe and estimate from the side (for example, transition matrices estimated from Dataset). From this perspective, the most faithful approach would be to use the (ideal) transition matrix, which, combined with selectivity and personalization, will produce (in simulations) social dynamics by estimating which we will come to that transition matrix, as estimated from Dataset. However, for the sake of simplicity, we build our analysis upon the estimated transition matrix, assuming that this matrix does not depend on selectivity and personalization factors.

### 6.3. Data noisiness

The ideal-typical behavior of the system under the presence of noise in the transition matrix remains the same. Thus, we concentrate on how $\alpha$ and $\beta$ affect the limiting behavior of the assortativity coefficient.

Our analysis reveals that for a given network topology, the maximal value of assortativity depends positively on both $\beta$ and $\alpha$ (see Figure 6). This result is intuitively clear. On the one hand, by increasing $\alpha$, we reduce the likelihood that like-minded agents will have different opinions after an interaction. On the other hand, higher values of $\beta$ amplify the probability of opinion adoption; thus, neighboring agents are more likely to espouse similar positions after interaction. The minimal disturbance (in the Euclidean metric) we should make with the transition matrix to achieve the acceptable value of the assortativity coefficient is $\alpha = 0.02$ and $\beta = 0$ (in the case of highly clustered networks). The resulting transition matrix

$$P_{1,:,:} = \begin{bmatrix} 0.995 & 0.005 \\ 0.952 & 0.048 \end{bmatrix}, P_{2,:,:} = \begin{bmatrix} 0.066 & 0.934 \\ 0.029 & 0.931 \end{bmatrix}$$

will be able to produce a sufficiently homophilic social system. However, this estimation does not work for low-clustered networks, for which one should take a more disturbed transition matrix (by establishing higher values of $\alpha$ and $\beta$).

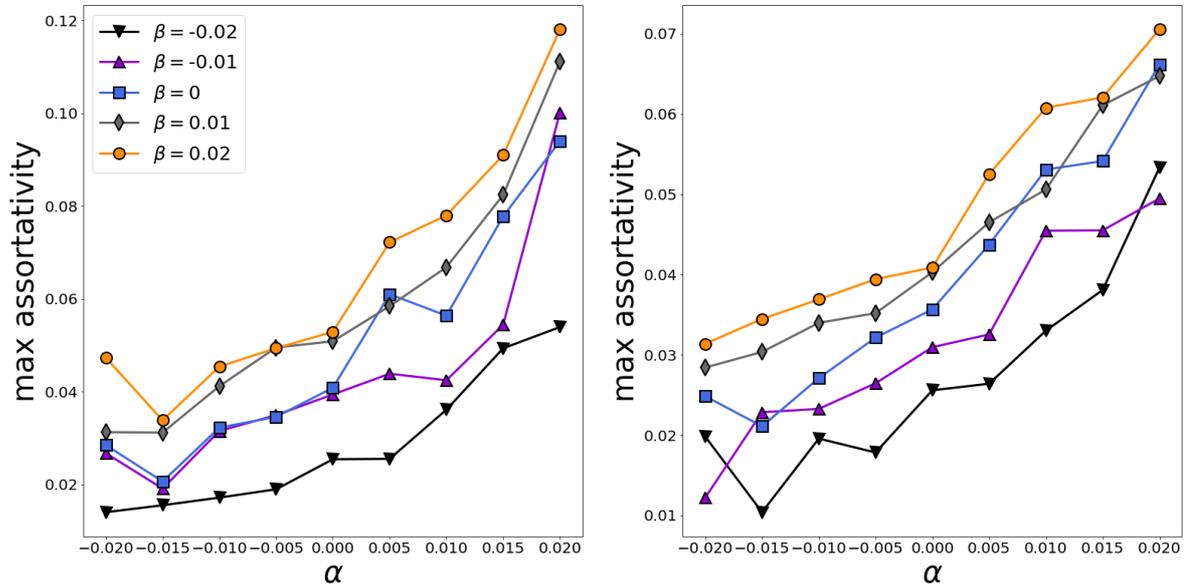

*Figure 6.* Maximal value of assortativity (during a single experiment) as a function of $\alpha$, separated by $\beta$ for two network topologies (left panel – random geometric, right panel – Erdős–Rényi).

## 6.4. Macroscopic behavior of Model 2

The presence of selectivity and personalization does not alter the qualitative behavior of the system (such behavior is inherited from Model 1). Besides, we have observed no situations when the social network becomes disconnected. Nonetheless, we notice that selectivity and personalization affect the limiting values of the macroscopic metrics (see Figure 7). Recall that our purpose is to determine the combination of $\gamma$ and $\delta$ that would increase the assortativity coefficient up to 0.14 at some point. Figure 6 indicates that higher selectivity rate values lead to more homophilic systems, as expected. However, personalization has the opposite effect on assortativity. Note that for large personalization rate values, the effect of selectivity is reduced. In limiting case $\delta \to 1$, we obtain the system characterized to be the zero level of assortativity regardless of the selectivity rate. In this case, only sufficiently similar opinions can interact; thus, all they can do is make antagonistic interactions. The most homophilic system is obtained if $\gamma = 1$ and there is no personalization; then, we get $C \approx 0.5$. Interestingly, the effect of topology observed for Model 1 (more clustered networks produce more homophilic systems) disappears when we increase the selectivity or personalization rates. Both selectivity and personalization have a positive effect on the system's polarization. However, in contrast to the assortativity coefficient, the dissimilarity coefficient varies in relatively small intervals $\approx [0.4, 0.51]$. As for assortativity, the effect of selectivity on polarization varies with the level of personalization: large values of the personalization rate discount the selectivity factor. In limiting case $\delta \to 1$, the dissimilarity coefficient fluctuates around the value $D \approx 0.51$ and does not depend on the selectivity rate.

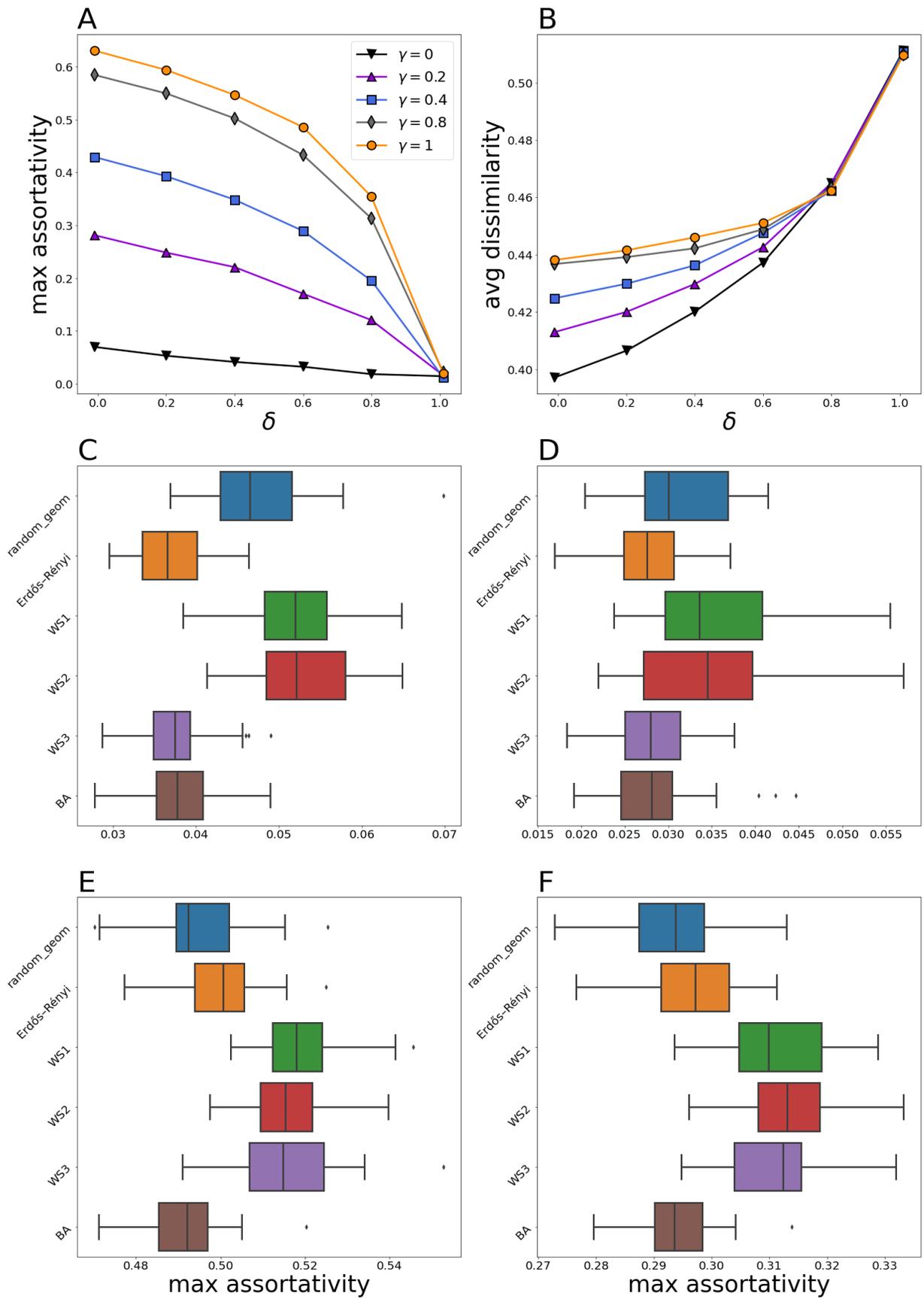

*Figure 7.* (Panels A, B). Maximal value of the assortativity coefficient and average (over time, starting from the moment the system had stabilized) value of the dissimilarity coefficient as functions of the personalization level, separated by the selectivity rate. The presented results are obtained on random geometric networks. (Panels C–F). The effect of topology on the assortativity coefficient for different combinations of selectivity and personalization levels: $\gamma = 0, \delta = 0$ (panel C), $\gamma = 0, \delta = 0.4$ (panel D), $\gamma = 0.6, \delta = 0$ (panel E), $\gamma = 1, \delta = 0.8$ (panel F). All results are derived for $m = 10$.

Based on these observations, we tune parameters $\gamma$ and $\delta$, attempting to get close to the empirical reference values in our simulations. The problem is that we need not only to achieve reference values but, what is important, do that *simultaneously*. We manage to resolve this problem in the triple opinion space by establishing $\gamma^* = 0.2, \delta^* = 0.1$. We begin again from opinion distribution $y_1(0) = 0.07, y_2(0) = 0.79, y_3(0) = 0.14$. In this case, the simulation system may reach the desired reference values at one point (see Figure 8). Furthermore, the local behavior of the macroscopic metrics near the reference values largely coincides with what we have noticed in the empirical data: the assortativity and dissimilarity increase, as does the fraction of individuals with the left-side opinions, whereas populations of individuals holding other (particularly middle-side) positions decrease. The asymptotic values of public opinion variables in this case are given by $y_1^* \approx 0.309, y_2^* \approx 0.573, y_3^* \approx 0.118$. They slightly differ from those obtained without selectivity and personalization (see Subsection 4.3). We do not state that there are no other combinations of $\gamma, \delta$, and configurations of initial opinions, which could also generate a system that is consistent with the empirics. For example, a little level of asynchrony is observed by selecting $\gamma = 0.2, \delta = 0$ or $\gamma = 0.2, \delta = 0.2$ (i.e., by varying the level of personalization). However, we find that changes in the level of assortativity will lead to an apparent discord (see Appendix).

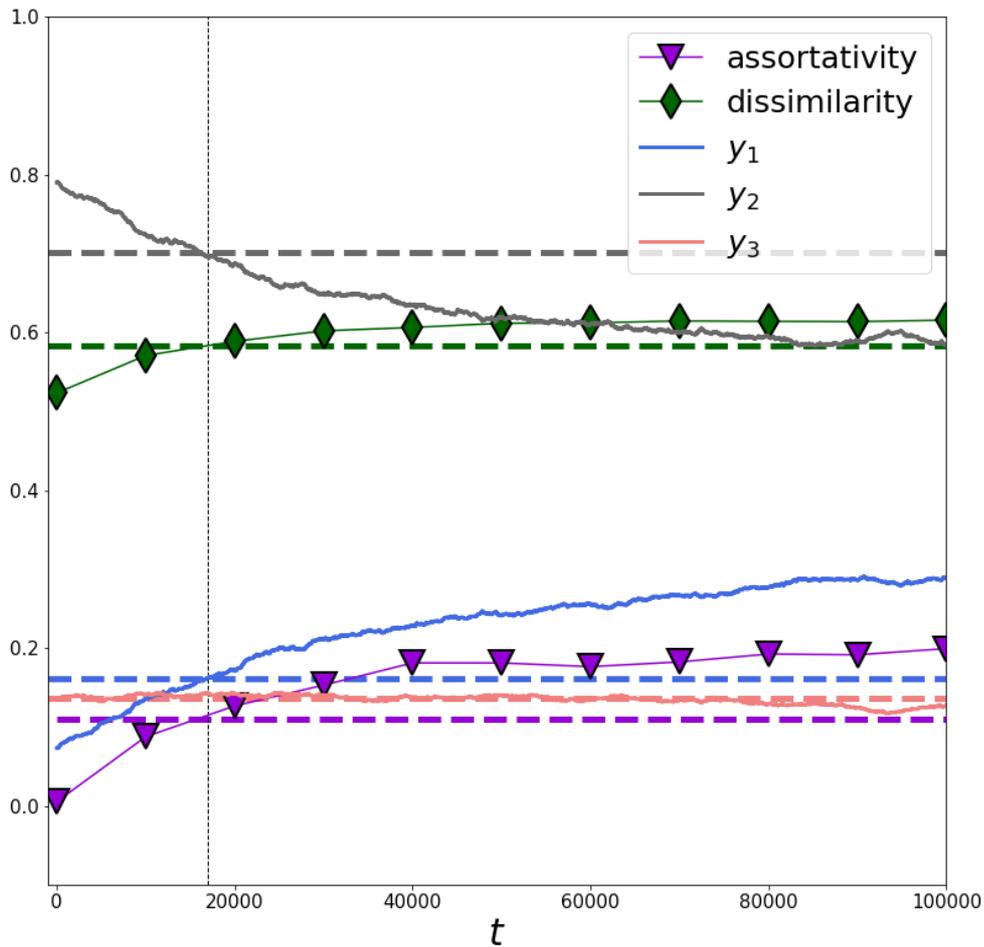

*Figure 8.* A simulation run for Model 2 if selectivity and personalization are $\gamma = 0.2, \delta = 0.1$ Initial opinions are drawn from distribution $y_1(0) = 0.07, y_2(0) = 0.79, y_3(0) = 0.14$. Dashed horizontal lines plot reference values from Table 2 (we take values from the first opinion snapshot). Common colors indicate the same metrics. The vertical dashed line marks the time point ($t \approx 17000$) at which all macroscopic metrics nearly coincide with the empirical reference values.

## 7. Discussion

Selectivity and personalization, which we implemented into the model, sufficiently advanced it and made it more realistic. Hence, we managed to simulate an artificial society that demonstrates properties similar to those observed empirically at the macro scale (at the particular point). Our finding on the (possibly) most appropriate settings that could generate empirically acceptable systems may be employed in several ways. On the one hand, these settings may offer an opportunity to understand how strong personalization and particularly selectivity are in the referenced OSN, from which the empirical data were gathered (see Appendix). More precisely, our results indicate that without selectivity, our artificial systems cannot achieve the desired empirics under the assumption that the transition matrix is estimated correctly. This knowledge can be further used in other studies in which this OSN is involved. However, this idea may be wrong in the case when the transition matrix is estimated with errors. Our current results do not answer the question of whether changes in the transition matrix could lead to *full* coincidence between simulations and empirics, but they at least hint that they *could do*. This problem requires additional analysis.

Next, knowing the current state of the system, we can predict its future evolution at the macro scale. However, only short-term predictions are meaningful because the transition matrix likely changes during long time intervals, reflecting events that occur both in this very OSN or beyond it. To be more specific, we calculate the transition matrix employing the second and third opinion snapshots from Dataset (recall that previous matrices were computed using the first two ones). We obtain the transition matrix

$$P_{1,:,:} = \begin{bmatrix} 0.976 & 0.024 & 0 \\ 0.967 & 0.033 & 0.000 \\ 0.94 & 0.058 & 0.002 \end{bmatrix}, P_{2,:,:} = \begin{bmatrix} 0.027 & 0.966 & 0.006 \\ 0.016 & 0.977 & 0.008 \\ 0.014 & 0.96 & 0.026 \end{bmatrix}, P_{3,:,:}$$

$$= \begin{bmatrix} 0.002 & 0.065 & 0.933 \\ 0.001 & 0.048 & 0.951 \\ 0 & 0.036 & 0.963 \end{bmatrix},$$

which slightly differs from its previous version (5). This matrix, implemented in Model 2 (with parameters $\gamma^* = 0.2, \delta^* = 0.1$ tuned above), marks new the equilibrium point $y_1^* \approx 0.338, y_2^* \approx 0.534, y_3^* \approx 0.128$, which is a more polarized opinion configuration compared to previous one ($y_1^* = 0.309, y_2^* = 0.573, y_3^* = 0.118$) obtained under transition matrix (5). From this perspective, the opinion dynamics of our model can be understood in terms of the evolution of the transition matrix.

## 8. Conclusion

In this paper, we introduced a minimal opinion formation model. A key ingredient of this model is the transition matrix that encodes how individuals with different opinions influence each other. Thus, the number of parameters in the transition matrix depends only on the number of possible opinion values rather than how many agents there are. We demonstrated that the elaborated model is quite flexible and can reproduce a broad variety of the existing micro-influence assumptions and models. At the same time, our model can be easily calibrated, provided that there is a dataset representing opinion dynamics processes. The proliferation of online social network platforms and advances made in the field of big data analysis ensure that such datasets will frequently appear. From this perspective, our model can be considered as a bridge, connecting theoretical studies on opinion formation models on the one hand and empirical research of social dynamics on the other.

We investigated the model analytically using mean-field approximation and numerically. We exemplified our analysis with the recently reported empirical data drawn from an online social network and demonstrated that the model can reproduce fragmented and polarizing social systems. In addition, we managed to simulate an artificial society that demonstrates properties qualitatively and quantitatively similar to those observed empirically at the macro scale. This result became possible after we had advanced the model with two important communication features: selectivity (the tendency of individuals to create ties with those with similar traits) and personalization algorithms, which decide what information a user will be exposed to, according to their previous online actions.

The model can be further advanced in several directions. First, it would be interesting to add in the model stubborn agents representing mass media or partisans. In its current form, the model cannot simulate a social network in which a few constantly stubborn agents are embedded. Further, the current form of OSN communications implies that users can make reposts or retweets and, thus, generate large-scale information cascades, which have an important impact on the information balance in the system (Fränken & Pilditch, 2021; Xie et al., 2021). A third possible direction concerns empirically analyzing the transition matrix's dynamics.

percolation and phase transitions of information on social media. *Nature Human Behaviour*, *5*(9), 1161–1168. https://doi.org/10.1038/s41562-021-01090-z46

# Appendix

## Appendix A. Mean-field approximation

Let us begin with some auxiliary facts. The probability of the opinion shift $x_s \to x_k$ after influence from an agent with opinion $x_l$ is the product of (i) the probability of selecting an agent with opinion $x_s$, (ii) the probability of choosing one of her peers with opinion $x_l$, and (iii) the quantity $p_{s,l,k}$. Because the social network is a complete graph, we obtain

$$\Pr[o_i(t+1) = x_k, o_i(t) = x_s, o_{i\leftarrow}(t) = x_l] = \frac{Y_s(t)}{N} \frac{Y_l(t) - \delta_{s,l}}{N} p_{s,l,k},$$

where $\delta_{s,l}$ is the Kronecker delta:

$$\delta_{s,l} = \begin{cases} 0, & s \neq l, \\ 1, & s = l. \end{cases}$$

As such, the probability of the occurrence of the shift that ends on opinion $x_k$ can be written as

$$\sum_{s=1}^{m} \sum_{l=1}^{m} \frac{Y_s(t)}{N} \frac{Y_l(t) - \delta_{s,l}}{N} p_{s,l,k}.$$

To obtain the probability of the shift that ends on opinion $x_k$ but does not begin on opinion $x_k$ (i.e., the likelihood that the population of agents holding opinion $x_k$ will increase by one), we should slightly modify the previous expression by adding the factor $1 - \delta_{s,k}$:

$$\Pr[Y_k(t+1) = Y_k(t) + 1] = \sum_{s=1}^{m} \sum_{l=1}^{m} \frac{Y_s(t)}{N} (1 - \delta_{s,k}) \frac{Y_l(t) - \delta_{s,l}}{N} p_{s,l,k}.$$

Let us rewrite this equation in a more complex form that nonetheless will be useful in the following computations:

$$\Pr[Y_f(t+1) = Y_f(t) + 1] = \sum_{s=1}^{m} \sum_{l=1}^{m} \sum_{k=1}^{m} \frac{Y_s(t)}{N} (1 - \delta_{s,f}) \frac{Y_l(t) - \delta_{s,l}}{N} p_{s,l,k} \delta_{k,f}.$$

Analogously, the chance that the population of agents having opinion $x_f$ will decrease by one (it may occur if and only if a randomly chosen agent with opinion $x_f$ will change it) is given by:

$$\Pr[Y_f(t+1) = Y_f(t) - 1] = \sum_{s=1}^{m}\sum_{l=1}^{m}\sum_{k=1}^{m} \frac{Y_s(t)}{N}\delta_{s,f}\frac{Y_l(t)-\delta_{s,l}}{N}p_{s,l,k}(1-\delta_{k,f}).$$

Given that we know agents' opinions at time $t$, the expectation of the number of agents with opinion $x_f$ at the next time step $t+1$ is given by:

$$E[Y_f(t+1)] = Y_f(t) + \Pr[Y_f(t+1) = Y_f(t)+1] - \Pr[Y_f(t+1) = Y_f(t)-1].$$

Substituting expressions for $\Pr[Y_f(\tau+1) = Y_f(\tau)+1]$ and $\Pr[Y_f(\tau+1) = Y_f(\tau)-1]$ into the previous formula, we get:

$$E[Y_f(t+1)] = Y_f(t) + \sum_{s,l,k} \frac{Y_s(t)}{N}\frac{Y_l(t)-\delta_{s,l}}{N}p_{s,l,k}\bigl[(1-\delta_{s,f})\delta_{k,f} - \delta_{s,f}(1-\delta_{k,f})\bigr]$$

or:

$$E[Y_f(t+1)] = Y_f(t) + \sum_{s,l,k} \frac{Y_s(t)}{N}\frac{Y_l(t)-\delta_{s,l}}{N}p_{s,l,k}(\delta_{k,f}-\delta_{s,f}).$$

Let us summarize our findings. If we know agents' opinions at time $t$, then their opinions at time $t+1$ are given by:

$$E[Y_f(t+1)] = Y_f(t) + \sum_{s,l,k} \frac{Y_s(t)}{N}\frac{Y_l(t)-\delta_{s,l}}{N}p_{s,l,k}(\delta_{k,f}-\delta_{s,f}), \quad f \in \{1,\dots,m\}.$$

For large values of $N$, we can state that:

$$\frac{Y_l(t)-\delta_{s,l}}{N} \approx \frac{Y_l(t)}{N}$$

and replace $E[Y_f(t+1)]$ with $Y_f(t+1)$:

$$Y_f(t+1) = Y_f(t) + \sum_{s,l,k} \frac{Y_s(t)}{N}\frac{Y_l(t)}{N}p_{s,l,k}(\delta_{k,f}-\delta_{s,f}).$$

Establishing scaled time $\tau = \frac{t}{N}$ and scaled time step $\delta\tau = \frac{1}{N}$, we get:

$$Y_f(\tau+\delta\tau) = Y_f(\tau) + \sum_{s,l,k} \frac{Y_s(\tau)}{N}\frac{Y_l(\tau)}{N}p_{s,l,k}(\delta_{k,f}-\delta_{s,f}).$$

Dividing both sides of the equation by $N$ and introducing the normalized quantity $y_s(\tau) = \frac{Y_s(\tau)}{N}$, we obtain:

$$\frac{y_f(\tau + \delta\tau) - y_f(\tau)}{\delta\tau} = \sum_{s,l,k} y_s(\tau) y_l(\tau) p_{s,l,k} (\delta_{k,f} - \delta_{s,f})$$

Because $N$ is large, then $\delta\tau \to 0$, and we finally come to the nonlinear autonomous system of differential equations:

$$\frac{dy_f(\tau)}{d\tau} = \sum_{s,l,k} y_s(\tau) y_l(\tau) p_{s,l,k} (\delta_{k,f} - \delta_{s,f}), \qquad f \in \{1, \dots, m\}.$$

*Appendix B. Dataset organization*

Dataset contains information on the dynamics of political preferences of a large-scale sample $I$ (approximately 1.6 M) of VKontakte (the most popular Russian social network) users. The sample was made by randomly choosing active (at least one platform interaction per month) individuals who meet a set of natural criteria (see (Kozitsin, 2020, 2021) for details). Further, the sample was additionally cleared of isolated subgroups of online friends in such a way that the resulting social network (whereby edges represent online friends) consists of one (giant) connected component. Note that only 0.7% of all nodes were removed during this procedure.

Users' opinions were estimated on a continuous opinion scale [0,1], where extreme positions 0 and 1 represent maximal opposition and support for the current Russian government correspondingly. Dataset is organized as three opinion snapshots made in February, July, and December 2018: $\hat{o}(t_1)$, $\hat{o}(t_2)$, and $\hat{o}(t_3)$, where the symbol "^" emphasizes that it provides only estimations of opinions. These estimations were made by applying a machine-learning methodology. Users' subscriptions to Vkontakte information sources were used as features, reflecting the fact that individuals tend to consume information coherent with their current opinions. As such, the dynamics of users' subscriptions to information sources should reflect the dynamics of their opinions.

The second component of Dataset is the symmetric adjacency matrix $\hat{A}$, the elements $\hat{a}_{ij} \in \{0,1\}$ of which represent friendship connections between the sample users at time point $t_2$. As was mentioned above, the corresponding social network is a connected graph.

As a result, each user $i$ from Dataset is characterized by the sequence of their own opinions $\hat{o}_i(t_1)$, $\hat{o}_i(t_2)$, and $\hat{o}_i(t_3)$ and by the sequence of average opinions of their friends $\hat{o}_{-i}(t_1)$, $\hat{o}_{-i}(t_2)$, and $\hat{o}_{-i}(t_3)$, where

$$\hat{o}_{-i}(t) = \frac{\sum_j \hat{a}_{ij} o_j(t)}{\sum_j \hat{a}_{ij}}$$

Note that Dataset is built under the assumption that friendship connections are static, representing one of its main disadvantages.

*Appendix C. Transition matrix estimation*

Let us now describe how we integrate Dataset information into the model. We will consider the case $m = 2$; other situations are elaborated analogously. We discretize the empirical opinion scale $[0,1]$ by endowing users who have opinion values from the interval $[0,0.5]$ with new opinions $x_1$ (say, $x_1 = -1$). Analogously, those who have opinions from the interval $[0.5,1]$ are marked with $x_2$ (say, $x_2 = 1$). A similar procedure is applied on average opinions of users' friends. Next, for each $s, l$, and $k$, we calculate quantity $p_{s,l,k}$ as follows:

$$p_{s,l,k} = \frac{\#\{i \in I \mid (\hat{o}_i(t_1) = x_s) \& (\hat{o}_{-i}(t_1) = x_l) \& (\hat{o}_i(t_2) = x_k)\}}{\#\{i \in I \mid (\hat{o}_i(t_1) = x_s) \& (\hat{o}_{-i}(t_1) = x_l)\}}, \quad (A1)$$

where $\#\{...\}$ denotes the cardinal number of the set. To put it simply, (A1) is the fraction of individuals who made opinion change $x_s \to x_k$ among those whose opinion is $x_s$ and whose friends' average opinion is equal to $x_l$. To avoid noise, we additionally require that the opinion shift should have a magnitude of more than 0.05 on the continuous scale. Note that in (A1), we first use two snapshots to calibrate the transition matrix's elements. Acting analogously, one can estimate the transition matrix from second and third opinion snapshots.

We should say that substantial opinion diversity, presented in Dataset, ensures that all combinations of $x_s$ and $x_l$ are available in sufficient quantity, providing the opportunity to estimate all the transition matrix elements (see Table C1).

Table C1

The joint distribution of users' opinions and opinions of their online friends (the first opinion snapshot, $m = 10$)

|  |  | The average opinion of online friends ||||||||||
|---|---|---|---|---|---|---|---|---|---|---|---|
|  |  | $x_1$ | $x_2$ | $x_3$ | $x_4$ | $x_5$ | $x_6$ | $x_7$ | $x_8$ | $x_9$ | $x_{10}$ |
| User's opinion | $x_1$ | 371 | 699 | 3,833 | 21,084 | 24,955 | 6,934 | 1,332 | 376 | 147 | 44 |
| | $x_2$ | 246 | 486 | 2,761 | 19,626 | 27,570 | 7,475 | 1,359 | 409 | 124 | 45 |
| | $x_3$ | 323 | 595 | 3,307 | 27,307 | 49,928 | 15,408 | 2,750 | 745 | 217 | 75.0 |
| | $x_4$ | 624 | 1,002 | 5,064 | 45,533 | 105,023 | 36,341 | 6,148 | 1,546 | 544 | 151 |
| | $x_5$ | 1,318 | 2,111 | 9,915 | 90,305 | 277,315 | 107,562 | 18,666 | 4,586 | 1,575 | 405 |
| | $x_6$ | 825 | 1,253 | 4,932 | 41,661 | 193,936 | 100,509 | 20,220 | 5,237 | 1,833 | 439 |
| | $x_7$ | 470 | 614 | 2,142 | 14,414 | 78,395 | 56,749 | 14,657 | 4,079 | 1,293 | 350 |
| | $x_8$ | 245 | 336 | 1,206 | 6,922 | 37,826 | 34,854 | 11,471 | 3,495 | 1,170 | 319 |
| | $x_9$ | 179 | 222 | 722 | 3,451 | 17,376 | 19,377 | 8,136 | 2,703 | 1,006 | 273 |
| | $x_{10}$ | 58 | 63 | 254 | 1,305 | 5,553 | 5,716 | 2,679 | 1,063 | 457 | 119 |

It is also important to note that our algorithm of the transition matrix estimation assumes many-to-one interactions (because we approximate the influence directed on a user by the average opinion of their friends), whereas our model is built on the idea of one-to-one interactions. One could suggest transforming the model and using the many-to-one assumption in simulations; however, we do not do so. The reason is rather technical—the averaging procedure may not be applicable for discrete opinion spaces.

## Appendix D. Tenfold transition matrix organization

Applying the algorithm presented in Appendix C, one can estimate the transition matrix in the case of the tenfold opinion space. In tables D1–D3, we present some slices of this matrix. In these tables, bolded columns indicate the probabilities of holding the current opinion (the most popular strategy observed). Importantly, the presence of both positive and negative influence can be found in these slices: if we increase $|x_s - x_l|$, then the probability that a user's opinion will be changed is raised, and changes both towards and outwards (see Table D2) influencing opinion become more likely. A more profound analysis of Dataset opinion dynamics can be found in Kozitsin (2020, 2021). The full version of the tenfold transition matrix can be found in Online Supplementary Materials.

Table D1

Slice $P_{1,:,:}$

| | | | | | | | | | |
|---|---|---|---|---|---|---|---|---|---|
| **0.942** | 0.038 | 0.003 | 0.009 | 0.009 | 0.000 | 0.000 | 0.000 | 0.000 | 0.0 |
| **0.938** | 0.044 | 0.005 | 0.006 | 0.005 | 0.000 | 0.003 | 0.000 | 0.000 | 0.0 |
| **0.945** | 0.042 | 0.007 | 0.003 | 0.002 | 0.000 | 0.000 | 0.000 | 0.000 | 0.0 |
| **0.947** | 0.037 | 0.008 | 0.004 | 0.003 | 0.001 | 0.000 | 0.000 | 0.000 | 0.0 |
| **0.939** | 0.036 | 0.009 | 0.006 | 0.005 | 0.003 | 0.001 | 0.001 | 0.000 | 0.0 |
| **0.924** | 0.043 | 0.009 | 0.008 | 0.007 | 0.004 | 0.002 | 0.001 | 0.001 | 0.0 |
| **0.925** | 0.045 | 0.009 | 0.006 | 0.008 | 0.002 | 0.002 | 0.002 | 0.000 | 0.0 |
| **0.908** | 0.065 | 0.006 | 0.003 | 0.000 | 0.009 | 0.003 | 0.006 | 0.000 | 0.0 |
| **0.896** | 0.067 | 0.037 | 0.000 | 0.000 | 0.000 | 0.000 | 0.000 | 0.000 | 0.0 |
| **0.881** | 0.048 | 0.000 | 0.000 | 0.000 | 0.000 | 0.048 | 0.024 | 0.000 | 0.0 |

Table D2

Slice $P_{5,:,:}$

| | | | | | | | | | |
|---|---|---|---|---|---|---|---|---|---|
| 0.002 | 0.001 | 0.011 | 0.080 | **0.854** | 0.051 | 0.002 | 0.000 | 0.000 | 0.000 |
| 0.001 | 0.004 | 0.009 | 0.067 | **0.873** | 0.040 | 0.004 | 0.001 | 0.001 | 0.000 |
| 0.001 | 0.004 | 0.010 | 0.070 | **0.874** | 0.038 | 0.003 | 0.000 | 0.000 | 0.000 |
| 0.001 | 0.002 | 0.008 | 0.060 | **0.896** | 0.031 | 0.001 | 0.000 | 0.000 | 0.000 |
| 0.001 | 0.001 | 0.005 | 0.054 | **0.895** | 0.041 | 0.002 | 0.001 | 0.000 | 0.000 |
| 0.000 | 0.001 | 0.005 | 0.059 | **0.871** | 0.059 | 0.004 | 0.000 | 0.000 | 0.000 |
| 0.001 | 0.001 | 0.005 | 0.068 | **0.837** | 0.078 | 0.008 | 0.001 | 0.000 | 0.000 |
| 0.000 | 0.002 | 0.008 | 0.068 | **0.827** | 0.083 | 0.011 | 0.003 | 0.000 | 0.000 |
| 0.000 | 0.001 | 0.005 | 0.066 | **0.826** | 0.090 | 0.009 | 0.003 | 0.000 | 0.001 |
| 0.000 | 0.003 | 0.013 | 0.090 | **0.790** | 0.092 | 0.005 | 0.005 | 0.000 | 0.003 |

Note: probabilities of negative shifts are denoted with lower font size numbers

Table D3

Slice $P_{10,:,:}$

| | | | | | | | | | |
|---|---|---|---|---|---|---|---|---|---|
| 0.000 | 0.0 | 0.000 | 0.000 | 0.000 | 0.000 | 0.018 | 0.000 | 0.125 | **0.857** |
| 0.000 | 0.0 | 0.000 | 0.000 | 0.000 | 0.018 | 0.018 | 0.000 | 0.161 | **0.804** |
| 0.000 | 0.0 | 0.005 | 0.000 | 0.005 | 0.005 | 0.000 | 0.014 | 0.096 | **0.876** |
| 0.001 | 0.0 | 0.000 | 0.000 | 0.006 | 0.005 | 0.005 | 0.005 | 0.081 | **0.897** |
| 0.000 | 0.0 | 0.000 | 0.001 | 0.002 | 0.004 | 0.004 | 0.005 | 0.064 | **0.919** |
| 0.001 | 0.0 | 0.000 | 0.001 | 0.001 | 0.002 | 0.005 | 0.005 | 0.076 | **0.909** |
| 0.000 | 0.0 | 0.000 | 0.000 | 0.002 | 0.002 | 0.006 | 0.007 | 0.079 | **0.904** |
| 0.001 | 0.0 | 0.001 | 0.000 | 0.000 | 0.003 | 0.006 | 0.005 | 0.074 | **0.910** |
| 0.000 | 0.0 | 0.000 | 0.000 | 0.005 | 0.000 | 0.002 | 0.014 | 0.079 | **0.900** |
| 0.000 | 0.0 | 0.000 | 0.000 | 0.000 | 0.000 | 0.000 | 0.018 | 0.071 | **0.912** |

*Appendix E. Influence of variations in selectivity and personalization on the system behavior*

Figures E1–E4 demonstrate that variations in $\delta^*$ and especially in $\gamma^*$ will lead to asynchronous behavior of the macroscopic metrics. More precisely, Figures E1–E4 plot simulation runs if selectivity and personalization are $\gamma = 0.2, \delta = 0$ (personalization is removed – see Figure E1), $\gamma = 0.2, \delta = 0.2$ (the level of personalization is increased – see Figure E2), $\gamma = 0.1, \delta = 0.1$ (the assortativity rate is decreased – see Figure E3), and $\gamma = 0.3, \delta = 0.1$ (the assortativity rate is increased – see Figure E4). Initial opinions are drawn from distribution $y_1(0) = 0.07, y_2(0) = 0.79, y_3(0) = 0.14$. Common colors represent the same metrics. Changes in the personalization level (Figures E1, E2) and, especially, in the assortativity rate (Figures E3, E4) response for making macroscopic metrics achieve reference values at different time points.

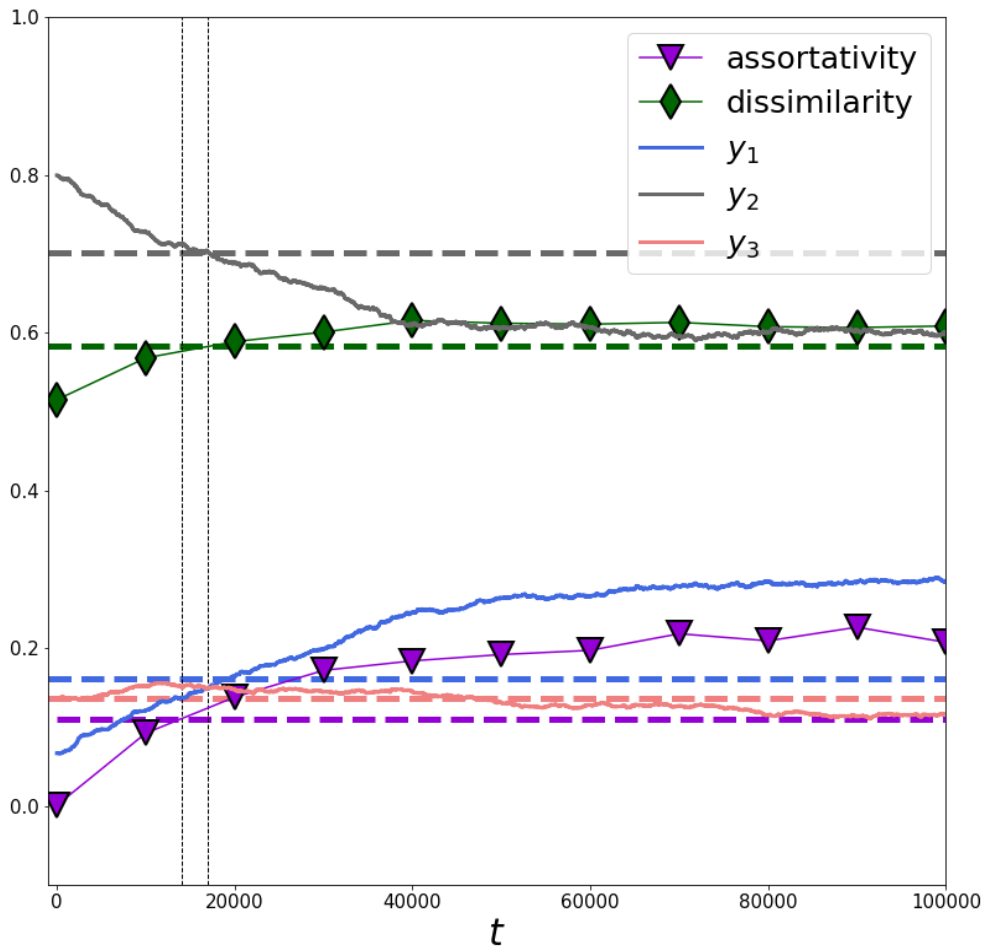

*Figure E1.* Macroscopic metrics achieve reference values with a little level of asynchrony (vertical dashed lines mark the time points at which macroscopic metrics coincide with the empirical reference values).

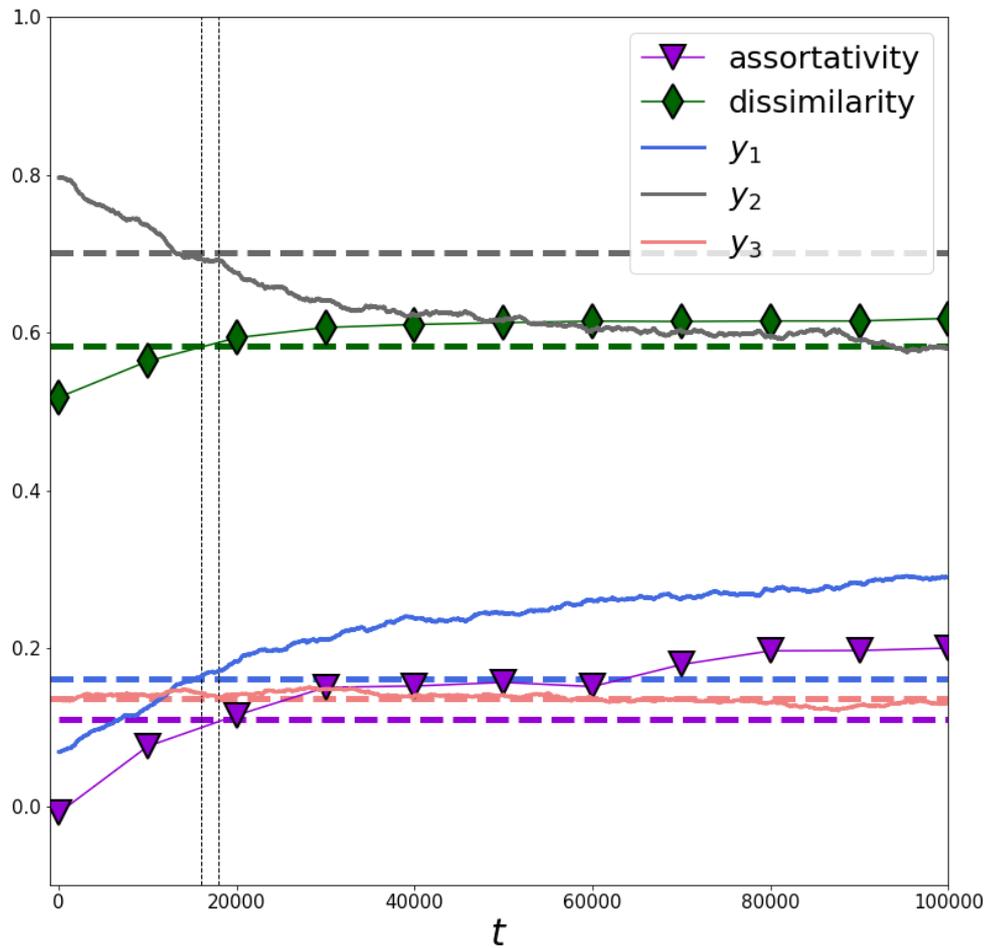

*Figure E2.* Macroscopic metrics achieve reference values with a little level of asynchrony (vertical dashed lines mark the time points at which macroscopic metrics coincide with the empirical reference values).

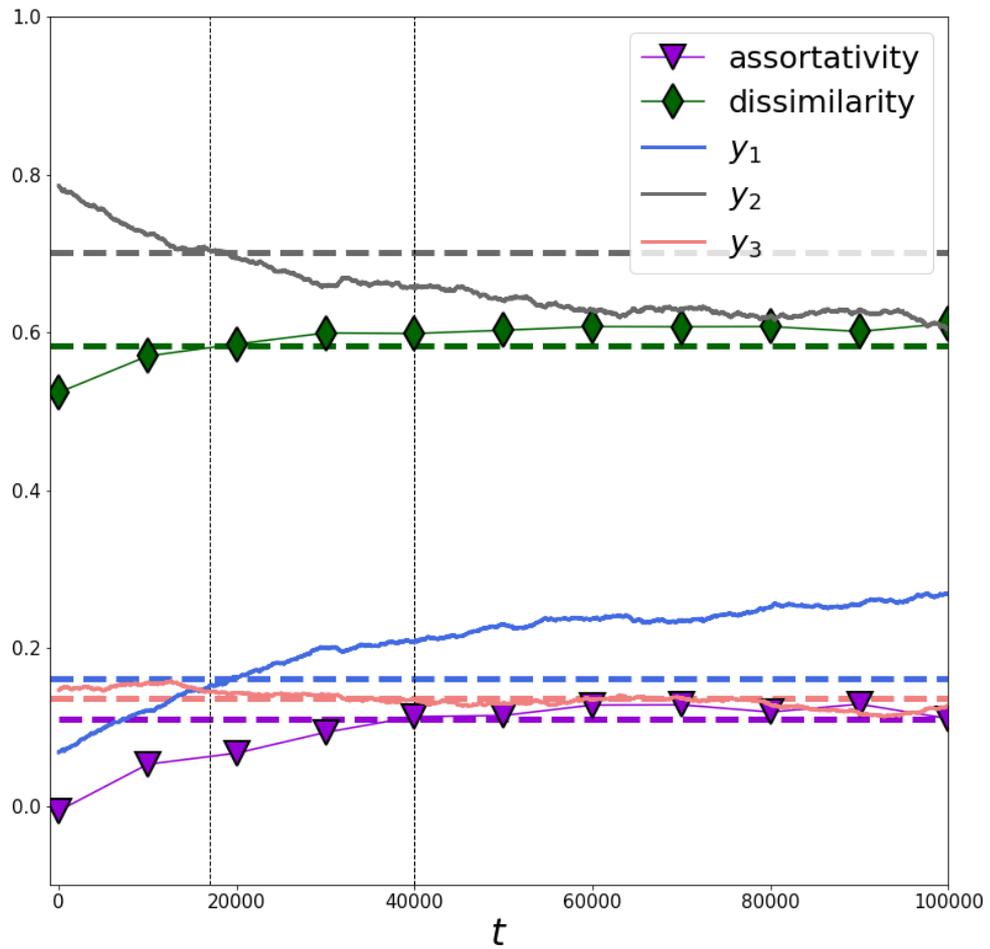

*Figure E3.* Macroscopic metrics achieve reference values at sufficiently different time points (vertical dashed lines mark the time points at which macroscopic metrics coincide with the empirical reference values).

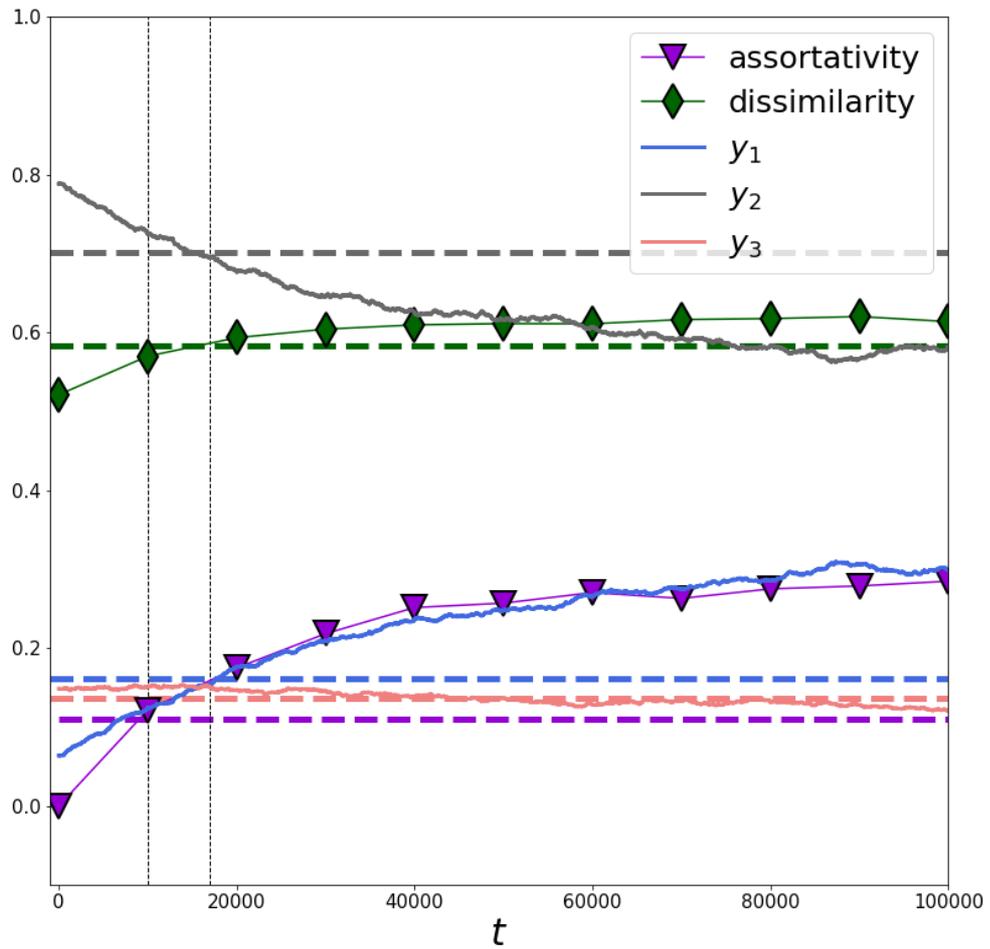

*Figure E4.* Macroscopic metrics achieve reference values at sufficiently different time points (vertical dashed lines mark the time points at which macroscopic metrics coincide with the empirical reference values).